\documentclass[]{aastex}
\usepackage{epsfig,emulateapj5}

\renewcommand{\sloppy}{}

\newcommand{\be}{\begin{equation}}
\newcommand{\ee}{\end{equation}}
\newcommand{\bt}{\begin{table} \begin{center}}
\newcommand{\et}{\end{center} \end{table}}
\newcommand{\ba}{\begin{eqnarray}}
\newcommand{\ea}{\end{eqnarray}}

\newcommand{\crho}{{\bar\chi}_\rho}
\newcommand{\cT}{{\bar\chi}_T}

\newcommand{\sch}{Schwarzchild}
\newcommand{\hz}{{\hat {\bf z}}}
\newcommand{\hx}{{\hat {\bf x}}}
\newcommand{\FE}{\tilde{E}}
\newcommand{\Frho}{\tilde{\rho}}

\newcommand{\Fp}{\tilde{p}}

\newcommand{\FS}{\tilde{S}}

\newcommand{\Fu}{\tilde u}
\newcommand{\Fw}{\tilde w}
\newcommand{\FF}{\tilde F}
\newcommand{\fac}{u_c}
\newcommand{\fack}{u_\kappa}
\newcommand{\facd}{u_\mr{d}}

\newcommand{\terma}{(k^{2}/3\chi_{v}\kappa_{v})}

\renewcommand{\u}{{\bf u}}
\def\dz{\frac{\partial}{\partial z}}

\newcommand{\Le}{{\cal L}_E}

\newcommand{\rhot}{\frac{\partial \rho}{\partial t}}

\newcommand{\kappaplus}{(\kappa + \sigma)}
\newcommand{\physr}{{Phys.~Rep.}}
\newcommand{\mr}{\mathrm}

\newlength{\figurewidth}
\newlength{\fullfigurewidth}
\newcommand{\citenp}[1]{\citeauthor{#1}~\citeyear{#1}}

\begin{document}

\submitted{To appear in ApJ, CITA-1999-35}

\title{ The Nature of the Radiative Hydrodynamic Instabilities
        in Radiatively Supported Thomson Atmospheres}

\author{Nir J. Shaviv$^{1,2}$}
\affil{1. Theoretical Astrophysics 130-33,
       California Institute of Technology, Pasadena, CA 91125 \\
       2. Current Address: Canadian Institute for Theoretical
       Astrophysics, University of Toronto \\ 60 St. George St.,
       Toronto, ON M5S 3H8, Canada}
\email{shaviv@cita.utoronto.ca}

\shorttitle{Radiative Instabilities}
\shortauthors{N.~J.~Shaviv}


\begin{abstract}

    Atmospheres having a significant radiative support are shown to be
    intrinsically unstable at luminosities above a critical fraction
    $\Gamma_\mr{crit} \approx 0.5-0.85$ of the Eddington limit,
    with the exact value depending on the boundary conditions.  Two
    different types of absolute radiation-hydrodynamic instabilities of
    acoustic waves are found to take place even in the electron
    scattering dominated limit.  Both instabilities grow over
    dynamical time scales and both operate on non radial modes.  One
    is stationary and arises only after the effects of the boundary
    conditions are taken into account, while the second is a
    propagating wave and is insensitive to the boundary conditions. 
    Although a significant wind can be generated by these
    instabilities even below the classical Eddington luminosity limit,
    quasi-stable configurations can exist beyond the Eddington limit
    due to the generally reduced effective opacity.

    The study is done using a rigorous numerical linear analysis of a
    gray plane parallel atmosphere under the Eddington approximation. 
    We also present more simplified analytical explanations.

\end{abstract}



\section{Introduction}

 Luminous objects shining close to the Eddington limit are of special
 interest since they occur in a wide range of astrophysical
 contexts. As such, they have been a popular topic for observational
 and theoretical investigations. Luminosities in the range of the
 Eddington limit appear in the very luminous stars populating the top
 of the HR diagram, in novae during the critical phases of the
 eruption and also in accreting objects of both galactic and
 extra-galactic scales. These very luminous objects usually display a
 great diversity of phenomena ranging from erratic variability to
 spatial structures and winds. Consequently, the theoretical modeling
 requires the incorporation of hydrodynamics, radiation and sometimes
 even magnetic fields.  Undoubtedly, much of the interesting behavior
 stems from the large, dynamically important luminosity of these
 objects -- luminosities that are close to the Eddington limit.

 The classical Eddington limit is the upper limit possible to the
 luminosity of an object in hydrostatic balance \citep{Edd}.  It is
 the luminosity for which the radiative force upwards balances the
 gravitational force downwards assuming the opacity of the (fully
 ionized) gas attains its minimal possible value, that is, the Thomson
 scattering opacity for non relativistic electrons.  For a homogeneous
 static configuration, it is the upper luminosity limit because above
 it there is no consistent solution to the hydrostatic equations.  An
 atmosphere that violates this upper limit will then simply blow
 itself apart -- an optically very thick wind is then unavoidable and
 the object will evaporate on a relatively short time scale.  When the
 opacity is larger than the Thomson scattering one, as is often the
 case, one can define a modified Eddington limit that is smaller than
 the `classical' value (e.g., \citenp{HD84}, \citenp{Lam}, and
 \citenp{App}).  The modified Eddington luminosity was used to explain
 the upper luminosity limit observed for stars and the variability
 that these stars exhibit.  The meaning of the modified Eddington
 luminosity should however be taken cautiously, because as one
 approaches the modified Eddington limit, the increased scale height
 and reduced density imply that in any fully ionized atmosphere, the
 opacity law is reduced to that of Thomson scattering \citep{HD94}. 
 Namely, as the modified limit is approached, it itself approaches the
 classical limit such that the modified Eddington limit is probably
 not an absolute luminosity limit.

 On top of the interesting questions on the nature of
 atmospheres when their luminosity approaches the Eddington luminosity,
 exists the fundamental question: `Is the Eddington luminosity the
 maximal luminosity under all conditions?'. Part of the answer was
 already answered in the negative by Shaviv (1998). Here we elaborate and
 extend the analysis and demonstrate that the classical Eddington
 limit is the limit under a very restrictive set of assumptions. Still
 we would like to emphasize that a significant factor in turning the
 Eddington limit into a major tool in setting astrophysical limits is
 the simplicity of the classical expression, namely $\Le=4 \pi c G M /
 \sigma_\mr{Thomson}$.

 There are several reasons why the study of radiative instabilities is
 important.  Most, if not all of the objects shining close to the
 Eddington limit show variability.  Novae for example, not only show
 very high (occasionally super-Eddington) luminosities over extended
 periods of time, they also show clumpy ejected material (e.g., the
 high resolution HST images, \citenp{Shar}) which could be explained
 using radiatively driven instabilities.  Another example are the
 super-Giants populating the top of the HR diagram.  Quite a few stars
 with ${M}\sim 100 {M}_{\sun}$ and ${L}\sim 10^{6}{L}_{\sun}$ exist
 across a wide range of spectral types and exhibit various degrees of
 variability.  For example, several of the numerous O3 type stars at
 30 Doradus are so bright that their luminosity is clearly very close
 to their Eddington limit (e.g., \citenp{MH}).  Less blue (and less
 young) are the Luminous Blue Variables (LBVs) termed so because of
 their often eruptive behavior in which their luminosity and rate of
 mass ejection can increase while their effective temperature
 decreases (see for example the review by \citenp{HD94}).  When these
 objects are in their eruptive state, at least some of them clearly
 surpass the Eddington limit for significant lengths of time.  For
 example, $\eta$ Car in its eruption between 1830 and 1860 released
 roughly $10^{49}~ergs$ of light over a period much longer than its
 dynamical time scale (e.g., \citenp{vGT}), yielding an average
 luminosity that is more than an order of magnitude higher than the
 Eddington limit (provided that the mass of $\eta$ Car is of the order
 of $100 M_{\odot}$).  The variability of these objects which is
 probably a consequence of the high luminosity, is therefore
 interesting to understand.

 An additional type of interesting objects are `very' and `super
 massive objects' (VMO's and SMO's) that might have existed as massive
 pop III stars in the young universe.  These massive stars are
 supposed to shine at luminosities very near the Eddington limit with
 the more massive of them at virtually the Eddington limit
 itself. Since the metalicity of these objects is extremely small, the
 analysis carried out here, which assumes that the opacity is
 dominated by Thomson scattering, is particularly relevant to these
 objects.

 Super-critical accretion disks are an altogether different type of
 relevant luminous objects.  Irrespective on whether they advect most
 of the excess accreted mass and energy above the Eddington rate into
 the black hole or whether the excess mass is blown away, each element
 of these disks should shine very close to local effective Eddington
 limit as these disks are always radiation pressure dominated.

 Various works have shown that atmospheres shining close to the
 Eddington limit exhibit a wide range of instabilities as well as
 other phenomena such as the generation of strong winds. For example,
 a large radiation pressure contribution reduces the adiabatic
 index. The relatively `loose' structure becomes unstable against
 convection when the adiabatic index decreases due to ionization
 (e.g., \citenp{Led}, \citenp{SC}, and \citenp{Wag}).

 \citet{JSO} analyzed the appearance of convection at high stellar 
 luminosities assuming that the condition for convective instability 
 is the classical {\sch} condition, namely the temperature gradient 
 must be higher than the adiabatic one.  The main concern of Joss et 
 al.~was the energy flux and its relation to the Eddington luminosity.  
 Here we are predominately interested in the {\em conditions} for the 
 stability under high radiative fluxes.  The main result of Joss 
 et al.~was that as the energy flux through an atmosphere is 
 increased, convection arises before the Eddington limit is 
 reached.  Moreover, the convective flux will always grow enough such 
 that the remaining radiative flux is less than Eddington, {\em as 
 long as convection is efficient}.  Namely, the interior of stars 
 which are dense, will never witness super Eddington fluxes.  It is 
 only near the surface where super Eddington fluxes can be reached.  Since 
 the optical depth below which convection becomes inefficient is relatively 
 high, the wind of a super Eddington object should be very thick and 
 the mass loss should be enormous if nothing happens above the 
 convective zone (cf \citenp{Sha00}).

 When the radiative force term becomes as important as the gravity or
 the gas pressure gradient terms, it can be a source of mechanical
 types of instabilities.  For example, \citet{GK} have shown that
 strange modes (which are `opacity' modified acoustic waves) suffer
 an instability in which mechanical work by the radiation is pumped
 into the waves.  This instability arises from the interaction of the
 acoustic waves with the non local nature of the diffusive radiation
 equations.  Another instability that requires special opacity laws is
 the $\kappa$-mechanism.  Unlike the $s$-mode instability, however, it
 grows on time scales much slower than the dynamical time scale, it
 does not involve work done by the radiation, and exists only when the
 system is neither fully adiabatic nor fully in the opposite (NAR)
 limit. However, it can exist when the system is far from Eddington, 
 because it involves the transfer of heat and not work by the radiation.

 A related instability arises when the radiation field transfers
 energy into acoustic waves due to high radiation pressures, as was
 shown by Hearn (1972, 1973\nocite{Hea72,Hea73}) and elaborated by
 \citet{BPR}, \citet{Sp}, \citet{Car} and \citet{Asp98}.  The
 instability arises when the opacity increases under compression and
 the change in the opacity is synchronized with the radiative force.
 This instability is not a real instability (aka `an absolute
 instability') in the sense that an infinitesimal perturbation will
 not grow to become nonlinear, however, as it is a secular drift
 instability (aka `a convective instability'), it can take finite
 amplitude waves (excited by convection for example) and increase
 their amplitude by many $e$-folds during one crossing time of the
 system.  (A counter propagating wave will be dissipated as fast, thus
 adding up to no net amplification of a standing wave unless the
 conditions for the strange mode instability are satisfied, e.g.,
 \citenp{G94}, \citenp{Pap97}).  An additional term arises when the
 Doppler variations of the radiative absorption (arising from wave
 motion) are taken into account \citep{Sha99}.  A third term is the
 driving through line absorption.  This term which utilizes the
 Doppler effect as well, is often very important in wind acceleration
 from hot stars (Owocki \& Rybicki 1984, 1985, 1986, 1991, \citenp{OR4},
 \citenp{FPP}).  The modes in systems with this type of simple drift
 instabilities cannot be used however to compose unstable standing
 waves in a simple Thomson scattering dominated atmosphere.

 The problem of the behavior of luminous atmospheres has also been
 compared to fluidized beds \citep{PS}.  Intuitively, the kind of
 bubbling observed in fluidized beds should also manifest itself in
 radiatively supported atmospheres.  Additional effects arise with the
 introduction of strong magnetic fields.  Specifically, when strong
 magnetic fields are present such that hydrodynamic motion
 perpendicular to the magnetic field lines is frozen, a linear
 analysis shows that `photon bubbles' arise at a high luminosity,
 facilitating the transfer of energy \citep{Aro}.  The original
 context of accretion onto magnetized neutron stars was also used to
 describe accretion disks as well \citep{Gam}.  Magnetic fields are
 neglected in the present work.  As we shall see, a plethora of
 instabilities can already manifest themselves when the magnetic
 fields are dynamically not important.

 In a recent work by \citet{ST99}, it was shown using a global
 non-radial linear mode analysis, that atmospheres can have unstable
 standing modes (as opposed to the secular drift instability, or
 convective instability) even in the case of pure Thomson scattering
 (as opposed to strange mode instability).  These unstable modes were
 shown to be horizontally propagating waves that amplify over a time
 scale much longer than the dynamical times scale of the system.  We
 have tried to recover this finding but could not succeed.  We report
 in the Appendix how such an instability could artificially arise. 
 Moreover, the two other soon be described instabilities are
 dynamically much more important.

 In addition to the intrinsic variability and the generation of winds,
 perhaps the most interesting consequence of the instabilities is the
 change induced in the Eddington limit. \citet{Sha98} has shown that
 in the presence of inhomogeneities, the spatially averaged flux
 through an atmosphere can increase without increasing the time
 averaged force on it. This arises because of the tendency of
 radiation to find the paths of least resistance when escaping through
 an atmosphere. Under some opacity laws, this will translate into
 having more radiation in regions where there is less mass and a
 smaller average force will be exerted. Thus, the luminosity can
 exceed the Eddington luminosity while the average force on a parcel
 of mass can remain less than the critical one.

 In summary, the classical Eddington limit is restricted to complete
 homogeneous systems and is frequently not valid in inhomogeneous
 ones.  A fundamental question of how such inhomogeneities are formed
 and what form they accept is discussed only briefly in Shaviv
 (1998). Here we expand on this question with emphasis on Thomson
 scattering dominated atmospheres (`Thomson atmospheres' hereafter).

 Evidently, a picture which relates all the different instabilities
 together and gives sound physical reasoning to their origin is mostly
 absent even though it is of paramount importance to the understanding
 of very luminous objects. It is therefore, our goal in this paper to
 address the following points:

\begin{enumerate}
\rightskip 0pt
  \item We wish to unravel all the absolute instabilities that Thomson
  atmospheres (which are the simplest of all) exhibit as they approach
  the Eddington limit.  This is done using a rigorous linear analysis
  (that is similar to but not the same as that of \citenp{ST99}) which
  assumes nothing save a gray slab atmosphere under the Eddington 
  approximation.
  
  \item We wish to understand the physical basis behind the
  instabilities found and how systems with these instabilities behave.

\end{enumerate}

 Two instabilities will be found and described.
 Additional instabilities arise  when the effects of a varying opacity
 are taken into account.   As the Eddington limit is
 approached, these instabilities generally become less important as
 the opacity becomes mostly that of scattering. However, a future 
 `tabulation' of all the instabilities of luminous atmospheres 
 should include them as well.

 A few clear extensions, such as a complete nonlinear analysis using
 hydrodynamic simulations or a systematic study of the parameter
 space, will be analyzed in forthcoming publications. These will also
 include a more elaborate global analysis of the non-trivial opacity
 law triggered instabilities.

 The paper is organized in the following way: The governing radiative
 hydrodynamic equations are presented in \S\ref{sec:equations}.  To
 clarify the physical picture, we discuss in \S\ref{sec:dimen} the
 parameter space of the problem and show where and when the various
 limits appear.  We then proceed in \S\ref{sec:fulllinear} to analyze
 the problem in a full global analysis assuming only a gray plane
 parallel atmosphere, calculate the eigenmodes, and find the
 characteristics of the instabilities found.  We continue in
 \S\ref{sec:TypeI} to describe the first instability and in
 \S\ref{sec:TypeII} the second instability.  We end with a discussion
 and summary of conclusions.  An appendix is given to relate the
 results found here to those found by \cite{ST99}.

\section{The radiative hydrodynamic equations and notation}
\label{sec:equations}

 We begin by laying out the equations used in this work. The equations
 describing the gas are the continuity, momentum and energy
 equations:
\ba
   \rhot + \nabla\cdot(\rho \u) &=& 0
\label{eq:contin}\\
   \rho {D \u \over D t} 
   &=& -\nabla p - \rho
   g \widehat{\bf z} + \rho \frac{\chi}{c} {\bf F}
\label{eq:moment}\\
    {D p \over D t} - c_{a}^{2} {D \rho \over D t}
    &=& -(\gamma-1)\rho \kappa c(S-E),
\label{eq:energy}
 \ea with standard notation: $p$, $\rho$ and $\u$ are the gas (matter)
 pressure, density and velocity.  $\chi$ is the total opacity per unit
 mass, which is the sum of the scattering $\sigma$ and the absorption
 $\kappa$ opacities.  Throughout this paper, opacity coefficients
 without subscripts denote opacity per unit mass while a subscript $v$
 denotes opacity per unit volume (with dimensions of $length^{-1}$). 
 $S$ is the equilibrium radiation energy density calculated from the
 gas temperature $S=4\sigma_\mr{SB} T^4/ c $ where $\sigma_\mr{SB}$ is
 the Stefan Boltzmann constant.  $E$ is the radiation energy density
 and ${\bf F}$ is the radiation energy flux.  The radiation energy and
 flux are those measured in the material rest frame, namely, the
 `Lagrangian' quantities.  $D /Dt \equiv \partial / \partial t + \u
 \cdot \nabla$ is the co-moving derivative.  We distinguish between
 the adiabatic speed of sound $c_{a}^{2}=\gamma p/ \rho$ and the
 isothermal speed of sound $c_{T}^{2}= p/ \rho$.  Gravity acts
 downward along the $\widehat{\bf z}$ direction.

 In view of the above assumptions, we describe the radiation field
 using the Eddington approximation. Its `Lagrangian' equations, valid 
 to ${\cal O}(v/c)$, are (\citenp{MWM}):
\ba
   { D E \over D t} - {4 E \over 3 \rho} {D \rho \over D t} + \nabla
   \cdot {\bf F } &=& \rho \kappa c (S - E)
\label{eq:radE}\\
   { D {\bf F }\over D t}+ \frac{c^2}{3} \nabla E &=&
   -\rho \kappaplus c {\bf F}.
\label{eq:radF}
\ea
  The second term in the first equation is the `compression' term which
  corrects the stationary frame radiation quantities into the material
  rest frame. This `Doppler' term describes radiation drag.
 
 The Eddington approximation is a reasonable assumption since we are
 interested in regions where $\infty > \tau \gtrsim 1$.  Even at the
 photosphere where the approximation is least accurate, it yields
 reasonable quantitative results and the correct qualitative results
 (e.g., \citenp{MWM}, pp. 357, 518).

 Once we apply a perturbation, quantities with an index `0' are the
 unperturbed values while an index `1' indicates the first order
 perturbation. The first order perturbation to the opacity,
 $\chi_{v,1}$, can be written as:
\be
   \chi_{v,1} = \chi_{v,0}\left[ \crho {\rho_1 \over \rho_0} + \cT
{T_1
\over T_0} \right],
\ee
 with the following definitions for the logarithmic derivatives:
\be
   \crho \equiv \left. \partial \ln\chi_{v} \over \partial \ln\rho
   \right|_T~~{\rm and}~~ \cT \equiv \left. \partial \ln\chi_v \over
   \partial \ln T \right|_\rho.
\ee
 The pure Thomson scattering case corresponds to $\crho=1$ and
 $\cT=0$. However, the approximation we will use in this paper is
\begin{equation}
 0~\ne~\kappa/\sigma \ll 1,
\end{equation}
 namely, the absorption is very small compared to the scattering, so
 that the derivatives are to a good approximation those of pure
 Thomson.  However, the absorption is not zero and the gas absorbs a
 small amount of radiation. This allows at times for the gas to
 equilibrate its temperature with the radiation, something that for a
 purely scattering atmosphere cannot take place.

 The last equation is the equation of state of a perfect gas relating
 the gas variables:
\be
p= {\rho k_{B} T \over A m_H},
\ee
 where $A$ is the mean atomic weight of the gas and $k_{B}$ the
 Boltzmann constant.

 As the radiation field may not be in equilibrium with the matter, the
 expression for the opacity can be very complicated and in principle
 the opacity should be calculated consistently with the radiation
 field.  In the present work we restrict the discussion to a gray
 opacity.

 The Fourier transform of a quantity $x$ is denoted by ${\tilde x}.$


\section{Dimensional analysis and critical parameters}
\label{sec:dimen}

 The driving forces on the system are the terms that appear on the
 r.h.s.~of eq.~(\ref{eq:moment}). In equilibrium, the radiative and
 pressure forces balance the pull of gravity and consequently  the
 r.h.s.~vanishes. The relative importance of each term can be related
 to the ratio between the radiation pressure and the gas pressure:
\be
\beta \equiv {p_{\mr{rad}}\over p} = {E_{0}/3 \over c_{T}^{2}\rho_{0}},
\label{eq:beta0}
\ee
 where $p\equiv p_{\mr{gas}}$.

 A second related parameter which determines the ratio between the
 radiative and gravitational forces is: \begin{equation} \Gamma = {F
 \chi \over c g}, \end{equation} where $F$ is the radiative flux and
 $\chi$ the opacity per unit mass. $\Gamma$ is often called the
 Eddington parameter. If $\chi$ is taken to be the Thomson scattering
 cross-section, then $\Gamma=1$ corresponds to the Eddington
 limit. For other opacities (which are generally larger if the gas is
 ionized) $\Gamma=1$ corresponds to the Modified Eddington limit.

 The parameters $\beta$ and $\Gamma$ are simply related to each other
 in Thomson atmospheres, for which the opacity per unit mass is
 constant.  From the Eddington approximation we have that
\be
 E_{0} = (2+3 \tau) {F_{0}\over c},
\ee
 where $\tau$ is the optical depth for extinction.  The decrease in
 the effective gravitational pull $(1-\Gamma) g$ leads to an increase
 in the density scale height of the atmosphere, namely,
\be
 l_{\rho}\equiv {c^{2}_{T}\over g(1-\Gamma)}.
\ee
 If the radiative force is constant and isothermality is assumed, then
 the puffed up scale height is constant with height. In the case of an
 atmosphere with an adiabatic gradient, $c_{T}$ should be replaced
 with $c_{a}$, in which case the scale height is not constant with
 height any more.  To demonstrate the relation between $\beta$ and
 $\Gamma$ we assume for simplicity an isothermal photosphere. In this
 approximation,
\ba
\rho_{0}(z) &=& \rho_{0}(z=0) \exp(-z/l_{\rho}); \nonumber \\ ~p_{0}(z)
& = & c_{T}^{2}
\rho_{0}(z=0) \exp(-z/l_{\rho}).
\ea
 The optical depth from $z=\infty$ to a given point at a depth of $z$
 is
\be
 \tau=\int^{\infty}_{z} \rho \chi dz = \chi l_{\rho} \rho_0(z=0) \exp
 (-z/l_{\rho}) = \chi l_{\rho} p_{0}(z) / c^{2}_{T}.
\label{eq:tau0}
\ee
 For further reference we also have $\tau=\chi_v l_\rho$.  We thus
 find from eqs.~(\ref{eq:beta0}) through (\ref{eq:tau0}) that
\ba
 \beta &\equiv& {p_{\mr{rad}}\over p} = \left( 2+3 \tau\right) {l_{\rho}
 F_{0}\chi\over 4c \tau c_{T}^{2}}={(2+3\tau) \over 3\tau}{\Gamma
 \over 1-\Gamma} \nonumber \\ &\approx& {\Gamma \over (1-\Gamma)}~{\rm
 for}~\tau\gg 1.
\label{eq:betaGamma}
\ea
 Conversely, for an optically thick atmosphere, we have that
\be
\Gamma \approx {\beta \over \beta + 1}.
\ee
 Atmospheres that have a small radiation to gas pressure ratio will
 therefore have a luminosity that is considerably smaller than the
 Eddington luminosity (given by $\Gamma=1$). When the gas and
 radiation pressures equate, we have that $\Gamma \approx 1/2$. When the
 radiation pressure is much larger than the gas pressure, $\Gamma$
 approaches unity (such that $1-\Gamma \ll 1$). In general, when the
 opacity is a function of height or when the non-isothermalness of the
 atmosphere is considered, we will get a more complicated relation
 between $\Gamma$ and $\beta$, however, we will still find that $\beta \ll
 1$ corresponds to $\Gamma \ll 1$, that $\beta\approx 1$ corresponds
 to $\Gamma \sim 1/2$ and that $\beta \gg 1$ corresponds to
 $(1-\Gamma)\ll 1$.

 Additionally important parameters arise when comparing the different
 time scales found in the problem. The first time scale is the dynamic
 time scale defined here as the isothermal sound crossing time of a
 density scale height:
\be
 \tau_\mr{dyn} \equiv {l_{\rho}\over c_{T}} = {c_{T}\over (1-\Gamma) g}.
\ee
 From analysis of equations (\ref{eq:radE}) and (\ref{eq:radF}), a
 second time scale appears.  To see this, we combine the two
 linearized versions of the equations assuming the heat absorption is
 negligible (i.e., in the adiabatic or isothermal limits) and
 neglecting the `Doppler' term, to get:
\be
 {1\over c^{2}}{\partial^{2} E \over \partial t^{2}} +{\chi_{v} \over
 c}{\partial E \over \partial t}-{1 \over 3} \nabla \cdot\nabla E = 0,
\ee
 where $\chi_{v}=\rho \sigma$. In the more general case the
 coefficient of ${\partial E / \partial t}$ is $\rho( \sigma + \kappa
 )/c$.  If the typical length scale is given by $k^{-1}$, a diffusion
 time scale can be defined as
\be
\tau_{\mr{diff}}(k)\equiv {3 \chi_{v} \over ck^{2}},
\ee
 when comparing the second and third terms.  Specifically, for a wavelength
 of the order of the scale height of the atmospheres, we can
 define:
\be
 \tau_{\mr{diff}}\equiv \tau_{\mr{diff}}(k=2 \pi l_{\rho}^{-1}) \equiv
 {3 \over (2 \pi)^2} {\chi_{v} l_{\rho}^{2}\over c}.
\ee
 A dimensionless parameter is its ratio to the dynamical time scale,
 given by
\ba
 r_{\mr{diff}}&\equiv& {\tau_{\mr{diff}}\over\tau_{\mr{dyn}}} =
 {3 \over (2 \pi)^2}{\chi_{v} l_{\rho}^{2} \over c} {c_{T}\over l_{\rho}}
 \\ \nonumber
 &=&
 {3 \over (2 \pi)^2}
 \left(c_{T}\over c\right) \left(\chi_{v}l_{\rho}\right) \equiv
 {3 \over (2 \pi)^2} \left(c_{T}\over c\right) \tau_{0},
\ea
 where $\tau_{0}\equiv \chi_{0} l_{\rho}$ is the typical optical depth
 of a density scale height. For $r_{\mr{diff}}\ll 1$, the radiation
 has ample time to converge to the diffusive solution given by the
 instantaneous mass configuration. In the opposite limit, the
 radiation does not have enough time to diffuse and any perturbation
 to it is synchronized with the gas.

 Note that for the optically thin part of the atmosphere, one should
 consider the nonlocal nature of the diffusive solution. In such a
 case, the diffusive time scale is simply given by the time it takes
 the radiation to traverse a scale height, i.e.,
 $\tau_{\mr{diff}}\approx l_\rho/c$, yielding $r_{\mr{diff}} \approx c_T/c$.

 A third time scale emerges from the energy equation
 (eq.~\ref{eq:energy}). We look at the set of
 eqs.~(\ref{eq:contin})-(\ref{eq:radF}) in the linearized case, when
 zero order gradients are neglected and when the terms arising from
 the finite propagation speed of light can be neglected as well.
 After applying a Fourier transformation, eqs.~(\ref{eq:radE}) \&
 (\ref{eq:radF}) become
\be
 {\FS}_{1}-{\FE}_{1} = {\FS}_{1}\left( \terma \over 1+\terma\right),
\ee
 where $\chi_{v}$ is the total extinction per unit volume and
 ${\FS}_{1}$ and ${\FE}_{1}$ are the Fourier transforms of $S_{1}$ and
 $E_{1}$ respectively.  We find that for $\terma \gg 1$,
 ${\FS}_{1}-{\FE}_{1}\approx {\FS}_{1}$ while for $\terma \ll 1$,
 ${\FS}_{1}-{\FE}_{1} \approx {\FS}_{1} \terma$. If we now plug this
 result into the linearized and Fourier transformed energy equation
 (eq.~\ref{eq:energy}) and use the first two hydrodynamic equations
 (eqs.~\ref{eq:contin} \& \ref{eq:moment}) to simplify it, we obtain
 after some algebra (in a similar manner to \S101 of \citenp{MWM})
 that
\be
 \left\{ \left[ \omega^2 - c_a k^2 \right] \omega + \tau_\mr{cool}^{-1}(k)
 \left[ \omega^2 - c_T k^2 \right] \right\} \Fp = 0,
\ee
where
\be
 \tau_{\mr{cool}}(k) \equiv {1 \over 12 \beta(\gamma-1) \kappa_v c
 }\left[1+ 3 {\chi_v \kappa_v \over k^2} \right].
\ee
 Clearly, when $\tau_{\mr{cool}}^{-1} \gg \omega$, the wave equation
 obtained will be the isothermal wave equation while for
 $\tau_{\mr{cool}}^{-1} \ll \omega$, the adiabatic wave equation is
 obtained. $\tau_{cool}(k)$ is therefore, the typical time scale which
 takes a wave with a wavevector $k$ to cool radiatively.

 We can define a dimensionless ratio between the cooling time scale
 for a wave with a wavelength of the order of the scale height and
 dynamical time scale that is the time it takes the same wave to
 cross a scale height as:
\ba
 r_{\mr{cool}} & \equiv & {\tau_{\mr{cool}}(k=2 \pi l_\rho^{-1}) /
 \tau_{\mr{dyn}} } \nonumber \\ &=& {c_T \over 12 \beta(\gamma-1) c \kappa_v l_\rho }
 \left(1+  {3 \chi_v \kappa_v l_\rho^2 \over (2 \pi)^2} \right)
\\  &\approx&
 \begin{array}{l l} \frac{1}{12 \beta(\gamma-1)} \frac{c_T}{c}
 \frac{1}{\tau_0}\frac{\chi_v}{\kappa_v} & {\rm for~~} \chi_v \kappa_v
 l_\rho^2 \ll 1 \\ \frac{1}{16 \pi^2 \beta(\gamma-1)} \frac{c_T}{c} \tau_0 &
 {\rm for~~} \chi_v \kappa_v l_\rho^2 \gg 1.  \end{array}
\ea
 For $r_{cool}\gg 1$, the cooling time scale is too long for the
 perturbations on scales of the scale height to have time to
 equilibrate to the radiation temperature. The gas is then close to
 the adiabatic limit. In the opposite limit of $r_{cool}\ll 1$, the
 gas has time to reach thermal equilibrium with the radiation and if
 the radiation has no perturbation, it is close to the isothermal
 limit. Since the equilibrium of the radiation can in principle be
 non-isothermal (if there is a net flux in the system), this limit is
 more generally called the NAR limit (non-adiabatic reversible), since
 in this opposite limit to adiabatic perturbations, the perturbations
 are again reversible.

\begin{figure*}[tbh]
\centerline{\epsfig{file=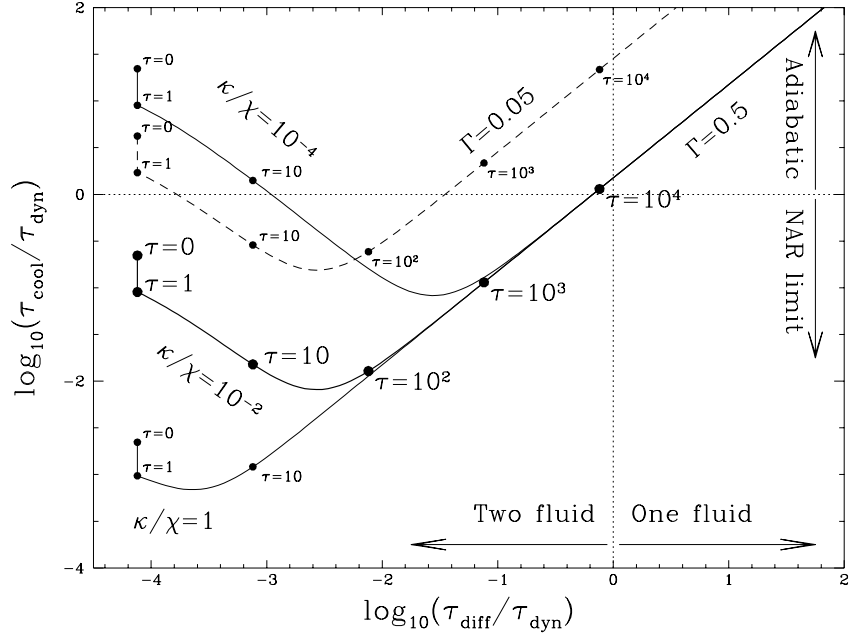,width=\fullfigurewidth,angle=-90}}
   \caption{ The connection between the ratio of the cooling time
   scale to the dynamical time scale and the ratio of the diffusion
   time scale to the dynamical time scale for a few sample Thomson
   atmospheres, as a function of optical depth.  The case depicted by
   a thick solid line corresponds to $\gamma =5/3$, $c_{T}/c=10^{-3}$,
   $\kappa/\chi=10^{-2}$ and $\Gamma=0.5$ ($\beta=1$ in the optically
   thick limit).  The two lighter solid lines correspond to changing
   the absorptive opacity from $\kappa / \chi = 10^{-2}$ to $1$ or to
   $10^{-4}$.  The dashed line corresponds to changing $\Gamma$ from
   $0.5$ to $0.05$.  The four regions correspond to whether the
   radiation is in the diffusion limit or not (i.e., whether it
   behaves as two fluids or one coupled fluid), and whether the gas is
   in the adiabatic or NAR limits.  Convection is effective in the top
   half.  The instabilities described here operate in the left half. 
   }
\label{fig:typical_ratios}
\end{figure*}

 Fig.~\ref{fig:typical_ratios} depicts the typical value of the
 dimensionless parameters in a typical Thomson atmosphere as a
 function of the optical depth from the top of it. Four distinct
 domains are seen in the $\tau_{\mr{cool}} / \tau_{\mr{dyn}}$ vs.
 $\tau_{\mr{diff}} / \tau_{\mr{dyn}}$ plane. The various
 approximations are depicted in this plane. The figure illustrates the
 run of the physical conditions in a typical star as one moves from
 the inside outwards, namely from high to low optical depths. The
 effect of the ratio of extinction to absorption on the path in the
 parameter plane is also shown.  The discussion in this paper
 concentrates on the l.h.s. of the parameter plane, that it, on the
 region in which the radiation does not evolve synchronously with the
 instantaneous gas configuration, i.e., when the radiation and gas
 cannot be considered as one fluid.

\section{Full global linear analysis}
\label{sec:fulllinear}

We begin with a complete global linear analysis. By changing the free 
parameters of the problem ($\Gamma$, boundary conditions, etc.) we 
expect to find the various instabilities that exist in this type of 
systems. Afterwards, we will analyze each one of them separately.

\subsection{Equations and Problem set-up}

 The governing set of equations is given in \S\ref{sec:equations}.
 After linearization and Fourier transforming the
 {\em horizontal} direction and {\em time} we get the following
 equations for the transformed quantities:
\begin{eqnarray}
 i\omega\frac{\Frho}{\rho_0} &=& ik \Fu + \dz \Fw + \Fw
 \frac{1}{\rho_0} \dz\rho_0
\\
 -i\omega\rho_0 \Fw &=& -\dz \Fp -g_* \Frho +\rho_0
 \frac{\kappaplus}{c} {\FF}_z
\\ \nonumber
 & & + \rho_0 \left(\kappaplus \over c\right) F_0 \left[ \chi_\rho
 {\Frho \over \rho_0} + \chi_p { \Fp \over p_0} \right]
\\
 -i\omega\rho_0 \Fu &=& -ik \Fp +\rho_0\frac{\kappaplus}{c} {\FF}_x
\\
 -i\omega \FE  +\dz {\FF}_z &=& \rho_0 \kappa c (\FS - \FE) - {4 E_0 \over 3 \rho_0} i \omega \Frho
 - ik{\FF}_x
\\
 -i\omega {\FF}_x + ik\frac{c^2}{3}\FE &=&-\rho_0 \kappaplus c {\FF}_x
\\
 -i\omega {\FF}_z + \frac{c^2}{3}\dz \FE &=& -\rho_0 \kappaplus c
 {\FF}_z - \Frho \kappaplus c F_0
\\ \nonumber
 & & - \rho_0 \kappaplus c F_0 \left[ \chi_\rho {\Frho \over \rho_0} +
 \chi_p {\Fp \over p_0} \right]
\\
 -i\omega \Fp - \rho_0 g_* \Fw &=& - c_s^2\rho_0( i k \Fu +\dz \Fw) \\
 & & -(\gamma-1)\rho_0\kappa c (\FS - \FE) \nonumber 
\\
 \FS &=& 4 E_0 (\frac{\Fp}{p_0} - \frac{\Frho}{\rho_0}),
\label{eq:perturbation}
\end{eqnarray}
 where $w$ is the vertical component of the velocity and $u$ the
 lateral one.  These equations are the equations that appear in
 \citep{ST99} with two differences.  First, we write the radiation in
 the co-moving system with the material.  Second, we retain the option
 of a variable opacity for generality and usage in the next work.  We
 omit for brevity the `$1$' index from the Fourier transformations
 of our eight variables $\Frho_1$, $\Fp_1$, $\Fu_1$, $\Fw_1$, $\FE_1$,
 $\FS_1$, ${\FF}_{x,1}$ and $\FF_{z,1}$.  These variables satisfy four
 complex algebraic equations and four complex ODE's with the height
 $z$ as the independent variable.

\noindent
 {\bf The Boundary Conditions:} The set of ODE's requires
  boundary conditions. Since the dependent variables
  $\Fp,\Fw, \FE, \FF_z$ are complex variables,
 four complex boundary conditions are required.

 Several gas boundary conditions can be imposed.  At either the bottom
 or the top we choose $w=0$ or $p=0$, which result with reflection of
 the waves at the boundary.  We find  that these conditions do not
 change the qualitative behavior of the waves. We should note however,
 that if we wish to compare our results to the
 behavior of real atmospheres, the results will have any meaning only
 if the waves in the real atmosphere considered can be trapped (and
 not for example propagate outward where they can be dissipated). We
 shall elaborate this point more in the discussion section.

 As for the radiation, there are more possibilities.

\noindent
 The following possibilities exist for the {\bf top}:

\noindent
\begin{tabular}{l l}
   Fixed radiation temperature: & $\FE(z_t)=0$ ~ or ~ \\
   Eddington $\tau=0$ condition: & $\FE(z_t)= 2 \FF(z_t)/c$
\end{tabular}

\noindent
While the following possibilities exist for the {\bf bottom}:

\noindent
\begin{tabular}{l l}
   Fixed radiation temperature: & $\FE(z_b)=0$ ~or~ \\
   Fixed incident flux: & $\FF(z_b)=0$ 
\end{tabular}

When the top of an atmosphere is free to emit radiation to infinity,
the Eddington condition for the top is the more appropriate
condition. The fixed radiation temperature will be the physically
appropriate one if for example the layer of interest is bounded at the
top by a layer that equilibrates quickly its temperature by
convection. As to the bottom conditions, they too can depend on the 
system at hand. For example, if the atmosphere is above a layer which 
conducts heat well, its temperature will be fixed. In principle, this 
boundary condition can be a mix of the two.

\subsection{Form of Solution}
\label{sec:solution}

In order to find the eigenmodes of the acoustic oscillations, we first
need to construct a profile of the unperturbed atmosphere.  We avoid
at this stage the usage of real stellar atmospheres since we are more
interested here in understanding the physical origin and behavior of
the unstable modes in the simple Thomson scattering dominated case and
not to quantify the unstable behavior of real stars that have complex
atmospheric opacities.  More realistic opacities as well as non LTE
effects will be included in future work.  As stated before, we also
neglect for simplicity the non-slab geometry of most relevant
atmospheres.

 The equations describing the unperturbed steady state atmosphere are
 the radiation diffusion equations:
\begin{eqnarray}
F_{0} & = & {\mathit const} \\ 
{d E_{0} \over d z} & = & -{3 \chi_{v,0} F_{0} \over c},
\end{eqnarray}
and the hydrostatic equation
\begin{equation}
{d p_{0} \over d z} = c_{T}^{2} {d \rho_{0} \over d z} = 
- \rho_{0} g \hz + {\chi_{v,0} F_{0} \over c}.
\end{equation}
 Since the gas and radiation are in thermal equilibrium in the
 unperturbed solution, we also have $T^{4}= cE/4 \sigma_{\mr{SB}}$
 which allows the calculation of $c_{T}$.

 Inasmuch as the Eddington relation should be satisfied in particular
 in the optically thin limit, we require that at the top of the
 atmosphere $E(\tau \rightarrow 0)= 2F/c$.

 The main parameter of interest in this work is $\Gamma = \chi_{v,0}
 F/c g$.  Hence, we solve for an unperturbed atmosphere with a given
 $\Gamma$ extending over a given optical depth range (say, between
 $\tau = 0.1-100$). The solution is carried out numerically and then
 tabulated so that a simple spline interpolation can be applied. The
 values of $p$, $\rho$ or $E$ can be quickly and to any given
 prescribed accuracy be found for any height in the atmosphere.

The next step is to find the eigenmodes of waves with a given
horizontal wavevector $k_{x}$ in a given atmosphere.  Our set of ODEs
has four first order complex ODE's and complex eigenvalues.  Two of
the boundary conditions are given at the bottom of the atmosphere
while the two others are given at the top, making it a complicated
boundary value problem.  To find an eigenmode we use the shooting
method.  The two `missing' bottom boundary conditions are iterated for
until the imposed top boundary conditions are satisfied.  The
procedure of finding the solution is carried out in the following way:

\begin{enumerate}
\rightskip 0pt
 \item Since two boundary conditions are given at the bottom, we guess
 the values of the two other variables that don't have a specified
 value on the boundary along with a guess for the unknown
 eigenfrequency $\omega$. For example, if $\Fp=0$ and $\FE=0$ are the
 bottom boundary conditions, we start by guessing the values of
 $\FF_z$, $\Fw$ and $\omega$.

 \item Once all the dependent variables are given or guessed on the
 lower boundary, we can integrate upwards.  The set of equations can
 be written as four ODE's and four algebraic equations that can be
 evaluated at each point $z$ in the following order:

\begin{itemize}
\rightskip 0pt
   \item Evaluate $\kappa, \sigma, \chi=\kappa+\sigma, E_0, s_\rho = -
   {\rho_0}^{-1} ({\partial \rho_0 / \partial z}) $ and $g_* = g -
   (\chi/c) F$.

   \item Calculate numerically the variables with algebraic equations:
\hskip -0.5cm
\begin{eqnarray}
\label{eq:numf}
 {\FF}_x & = & {k c^2 \FE \over 3(\fac \omega + i \rho_0 \chi c)}
\\
 \Fu &=& {k \Fp \over \omega \rho_0} + {i \over \omega} {\chi \over c}
 {\FF}_x
\\
 \FS & = & \left[  i\omega \Fp - c_s^2 \rho_0( i \omega
 {\Fp \over p_0} + s_\rho \Fw ) + \FE (\gamma-1) \rho_0 c\kappa_0 \fack^{-1} \right.
\nonumber \\  & &
\left. + \rho_0 g_* \Fw \right]
 / \left[ (\gamma-1) \rho_0 c \kappa_0 /\fack - {i \omega {c_s}^2 / 4 E_0} \right]
\\
 \Frho & = & \rho_0 \left( {\Fp\over p_0} - {\FS\over 4 E_0} \right).
\end{eqnarray}

  \item Integrate upwards using the following ODE's:
\begin{eqnarray}
   \dz \Fw &=& i \omega {\Frho \over \rho_0} - i k \Fu + \Fw s_\rho
\\
   \dz \Fp &=& i \omega \rho_0 \Fw - g_* \Frho + \rho_0 {\chi\over c}
   {\FF}_z \nonumber \\ &&+ \rho_0 F_0 {\chi \over c} \left[ \chi_{\rho }{\Frho \over
   \rho_0} + \chi_p {\Fp \over p_0} \right]
\\ \dz {\FF}_z &=& \fack \rho_0
   \kappa c (\FS - \FE) + \fac i \omega \FE - i k {\FF}_x - i \omega {4 E_0 \over 3 \rho_0} \facd
\nonumber \\ &&
\\ \dz \FE &
   = & - 3 \rho_0 {\chi\over c} {\FF}_z + {\fac 3 i \omega \over c^2}
   {\FF}_z - 3 \Frho {\chi\over c} F_0 \nonumber \\ && - 3 \rho_0 {\chi\over c} F_0
   \left[ \chi_{\rho} {\Frho \over \rho_0} + \chi_p {\Fp\over p_0}
   \right].
\label{eq:numl}
\end{eqnarray}
   We marked several terms so as to be able to trace their effect.
   The terms multiplied by $\fac$ are terms that originate from the
   wavy behavior of the radiation field, while terms multiplied by
   $\fack$ are terms that originate from the finite cooling time of
   the system.  The terms with $\facd$ are the `Doppler' terms.  In
   the normal case $\fac=\fack=\facd=1$.  In the limit
   $\fac\rightarrow 0$ we get the instantaneous limit for the
   radiation.  While in the limit $\fack\rightarrow 0$ we recover the
   negligibly small gas heat capacity limit (the NAR limit).  The
   limit $\facd \rightarrow 0$ shuts off the effects of the Doppler
   correction in the radiation field to the motion of the material.
   Since one of the goals is to check whether these terms are important for
   instability, they are specifically marked and can be
   artificially changed, though they are equal to unity in real cases.

   The numerical integration is accomplished using the fourth order
   Runge-Kutta integration scheme.
\end{itemize}

 \item At the top, the solution is compared with the two top boundary
 conditions. The initial guess for the two additional bottom
 conditions and the eigenfrequency $\omega$ are then corrected and the
 equations are integrated upwards once again. The `shooting' upwards
 it iterated until the solution converges to the required top boundary
 conditions. The solution obtained is an eigenmode of the system and
 its eigenvalue is the converged frequency $\omega.$
\end{enumerate}


 Since we are interested in finding more than one particular eigenmode,
 we use the above procedure to search for more modes. This search is
 performed by systematically guessing different initial guesses for
 the eigenvalue $\omega$. In this way, the entire frequency domain is
 covered up to a given frequency.

 In principle, other types of modes such as radiation diffusion modes
 at very high frequency can be searched for and found, however, they
 are beyond the interest of this work as they are generally very
 stable (with large damping rates, e.g.~\citenp{MWM}).  Similarly, we
 are not interested in convection or $g$-modes.  For any reasonable
 absorptive opacity larger than $\sim (c_{a}/c)(p_\mr{rad}/p_\mr{gas})
 \sigma$, the top part of the atmosphere will cool too quickly for
 convection to be efficient.  This will appear in the form of modes
 that have are real ($g$-modes) or pure imaginary (convective modes)
 with absolute oscillation or growth rates that are much smaller than
 the dynamical time scale.  Only if we force the atmosphere to be
 adiabatic by artificially having a very small absorptive opacity,
 will the $g$-modes or convective modes obtain a `dynamic' time scale.

 Once the eigenmodes of waves with a given $k_{x}$ and $\Gamma$ are
 found, we can systematically search for the eigenmodes of other
 $k_{x}$, or of atmospheres with a different $\Gamma$. Since we are
 interested in the locus of the eigenvalues as a function of either
 $k_{x}$ or $\Gamma$, we can evaluate the locus by changing $k_{x}$ or
 $\Gamma$ by small increments and use the previous value of the two
 independent variables and eigenvalue as the new initial guess.  We
 assume that the eigenvalues are continuous functions of $k_{x}$ and
 $\Gamma$. This allows a relatively fast construction of the locus of
 eigenvalues.

\subsection{Results}

 We divide the description of the results into three parts.  We first
 begin by analyzing the eigenvalues $\omega$ of an atmosphere with a
 given $\Gamma$ and horizontal wavenumber $k_{x}$ and summarize the
 two instabilities found.  We proceed to construct the eigenvalue
 spectrum of an atmosphere as a function of the horizontal wavevector
 $k_{x}$ and then perform a similar analysis to find the eigenvalues
 as a function of the Eddington parameter $\Gamma$.  In subsequent
 sections, we will then study the two absolute instabilities.

\subsubsection{An atmosphere with a given $\Gamma$ and $k_{x}$ and
 the instabilities found}

 As described in \S\ref{sec:solution}, for a given set of boundary
 conditions imposed, the eigenvalues corresponding to a given $\Gamma,
 k_{x}$ pair are found.

\begin{figure*}[tbh]
\centerline{\epsfig{file=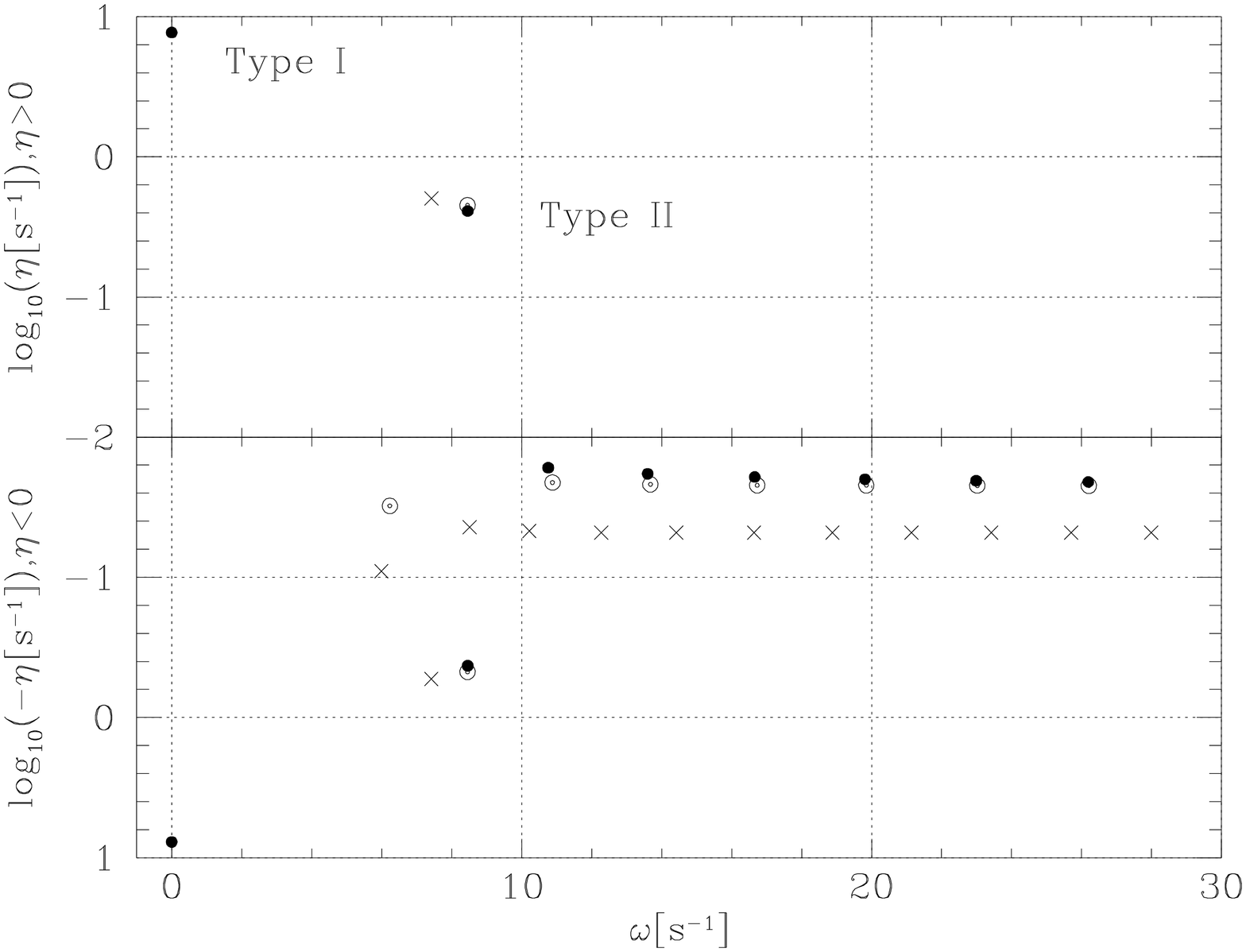,width=\figurewidth,angle=-90}}
\caption{The eigen frequencies of the lowest modes in three
atmospheres.  The two panels refer to the two possibilities of the
sign of $\eta$ -- the growth or damping rate of the modes.  The filled
circles are the modes of an atmosphere with $\Gamma=0.9$,
$k_x=1/l_{\rho}$, $g=10^8{cm~s^{-2}}$, $\kappa/\chi=1/100$ and a fixed
temperature imposed at a bottom placed at $\tau_{b}=10$ while the top
is at $\tau=0.3$.  The empty circles are the modes of the same
atmosphere with a fixed flux imposed at the bottom.  The crosses are
the eigenmodes obtained when the bottom boundary of the second
atmosphere is moved to $\tau=30$.  The Type I instability is seen in
only the first atmosphere with a fixed temperature at the bottom.  The
Type II instability is seen in all cases.}
\label{fig:atmos123}
\end{figure*}

 Fig.~\ref{fig:atmos123} depicts the lowest acoustic eigenvalues
 obtained for three atmospheres with $\Gamma=0.9$ and $k_{x}=1/
 l_{\rho}$.  One atmosphere has a fixed temperature imposed at the
 bottom boundary condition while the other two have a fixed flux
 imposed at the bottom.  In all three cases the Eddington
 approximation condition is imposed at the top (namely,
 $E_{t}=2F_{t}/c$), at $\tau_{t}=0.3$.  In the fixed temperature case,
 the bottom is fixed at $\tau_{b}=10$.  In the two fixed flux
 atmospheres, the bottom is located at $\tau_{b}=10$ and $30$.
 
 We assume that there is a small absorptive opacity $\kappa=\chi/100$.
 This absorptive opacity, however small, is still 100 times larger
 than the value needed to force the perturbations with a period of the
 dynamical time scale to be in the NAR limit and not the adiabatic
 one.  This is a better approximation for stellar atmospheres that
 most often have a small but significant absorptive opacity in this
 respect.  One the other hand, the absorptive opacity is too small to
 change the total opacity which is dominated by electron scattering.
 The gravitational acceleration in this particular case is $10^8
 cm~s^{-2}, $ an acceleration which corresponds to a White Dwarf
 envelope.  The physics obtained is to a large extent not a function
 of this parameter.  Choosing a different gravitational constant will
 just imply that the typical time scale will be different and the
 effective temperature needed to sustain a given $\Gamma$ is different
 as well.  In this particular case, the effective temperature obtained
 is $5.82\times 10^{5}{~}^\circ{\rm K}$.  At this temperature, the
 ratio $c_T/c$ at the center of the layer is roughly $1.0\times
 10^{-4}$.  (In reality these conditions are similar to those found in
 novae.)

 We can now see the two main instabilities.  When the lower boundary
 condition has a fixed temperature, the lowest vertical mode can be
 unstable to the first type of instability -- `Type I'. When unstable,
 this mode's eigenfrequency becomes purely imaginary. It grows on
 dynamical time scales while its counterpart decays with the same
 rate. The growth rate found in this case is $7.7 s^{-1}$. This is the
 dynamical time scale of the system.

 Under the two different lower boundary conditions, another type of
 instability can operate -- `Type II'.  When a mode succumbs to this
 instability, is has a real part with a somewhat smaller but
 significant positive imaginary part.  Namely, it is mostly traveling
 but it amplifies (and its counterpart decays) on a dynamical time
 scale.  In the case depicted here $\omega = (8 + 0.45i)~s^{-1}$ with
 only small changes between the three cases.  Although the growth rate
 of this mode is not as fast as fast as that of the Type I
 instability, the $e$-folding growth time of this mode is nevertheless
 only three oscillation periods.  Moreover, unlike the Type I
 instability, this one is relatively insensitive to the boundary
 conditions.
 
\begin{figure*}[tbh]
\centerline{\epsfig{file=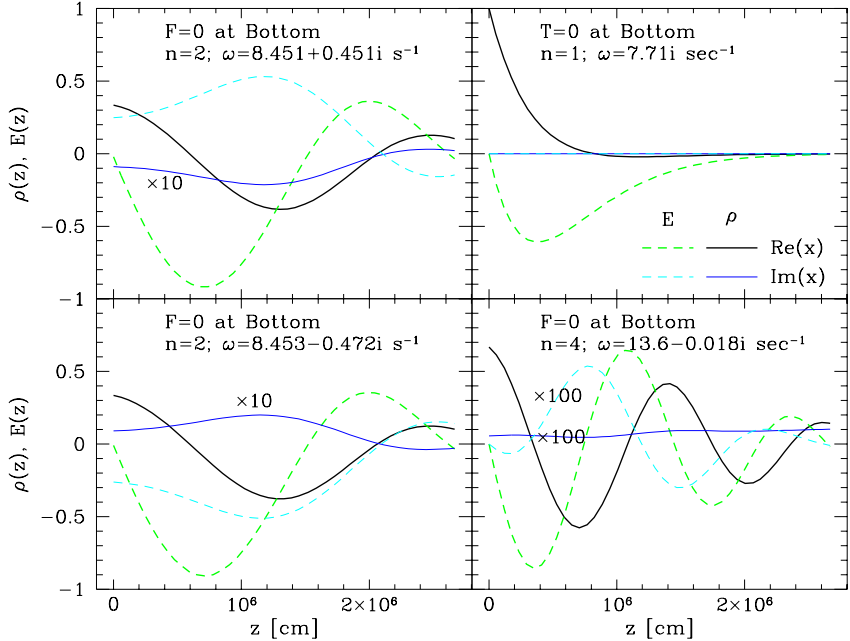,width=\fullfigurewidth,angle=-90}}
\caption{The profiles of unstable and stable modes.  The top left
panel describes the Type II unstable mode of the second atmosphere of
the previous figure (fixed flux at the bottom).  The bottom left panel
depicts the profile of the conjugate mode which is highly damped.  The
top right panel shows the profile of the Type I unstable mode of the
first atmosphere described in the previous figure (fixed $T$ at the
bottom).  The bottom right panel shows the profile of the fourth mode
in the same atmosphere.  It is slightly radiation damped.  The
variables plotted are the energy density $\FE$ (dashed lines) and the
matter density $\Frho$ (solid lines) multiplied by $\beta c_T^2$, so
as to be on the same par as $\FE$.  Since the problem is linear, the
absolute normalization of the modes is meaningless.  The real part is
given by heavy line while the imaginary part by a thin one.}
\label{fig:profiles}
\end{figure*}

 Fig.~\ref{fig:profiles} shows the structure of both unstable and stable
 modes. The main features seen are that the Type I mode
 has gas and radiation energy densities that are inversely correlated
 with each other. In the Type II mode, there is clearly a phase lag
 between the density and the radiation. This lag flips sign in the
 conjugate pair of the Type II unstable mode.
 
 We also checked whether the various artificial factors that we
 introduced change the occurrence of the modes.  $\fac$ changes the
 ratio between the speed of light and speed of sound.  We found that
 both instabilities exist in the two fluid limit (in which the
 radiation speed is infinite) but disappear when in the one fluid
 limit, when the radiation cannot diffuse.  $\fack$ changes the
 importance of the absorption opacity which determines the ratio
 between the cooling time scale and the dynamical time scale.  We
 found that both instabilities exist in both limits, the adiabatic
 limit in which the gas has no time to equilibrate with the radiation
 temperature and in the NAR limit, in which the gas temperature is set
 by the radiation temperature.

 We also searched for a third instability like the one described by
 \cite{ST99} but could not find it.  We defer further discussion on
 it to the appendix where we will try to understand how such an
 instability could arise artificially.  We also explain why the two
 dynamic instabilities found here were not found by Spiegel and Tao
 even though they did solve the same equations.
 
\subsubsection{Eigenmodes as a function of $k_{x}$}

\begin{figure*}[tbh]
\centerline{\epsfig{file=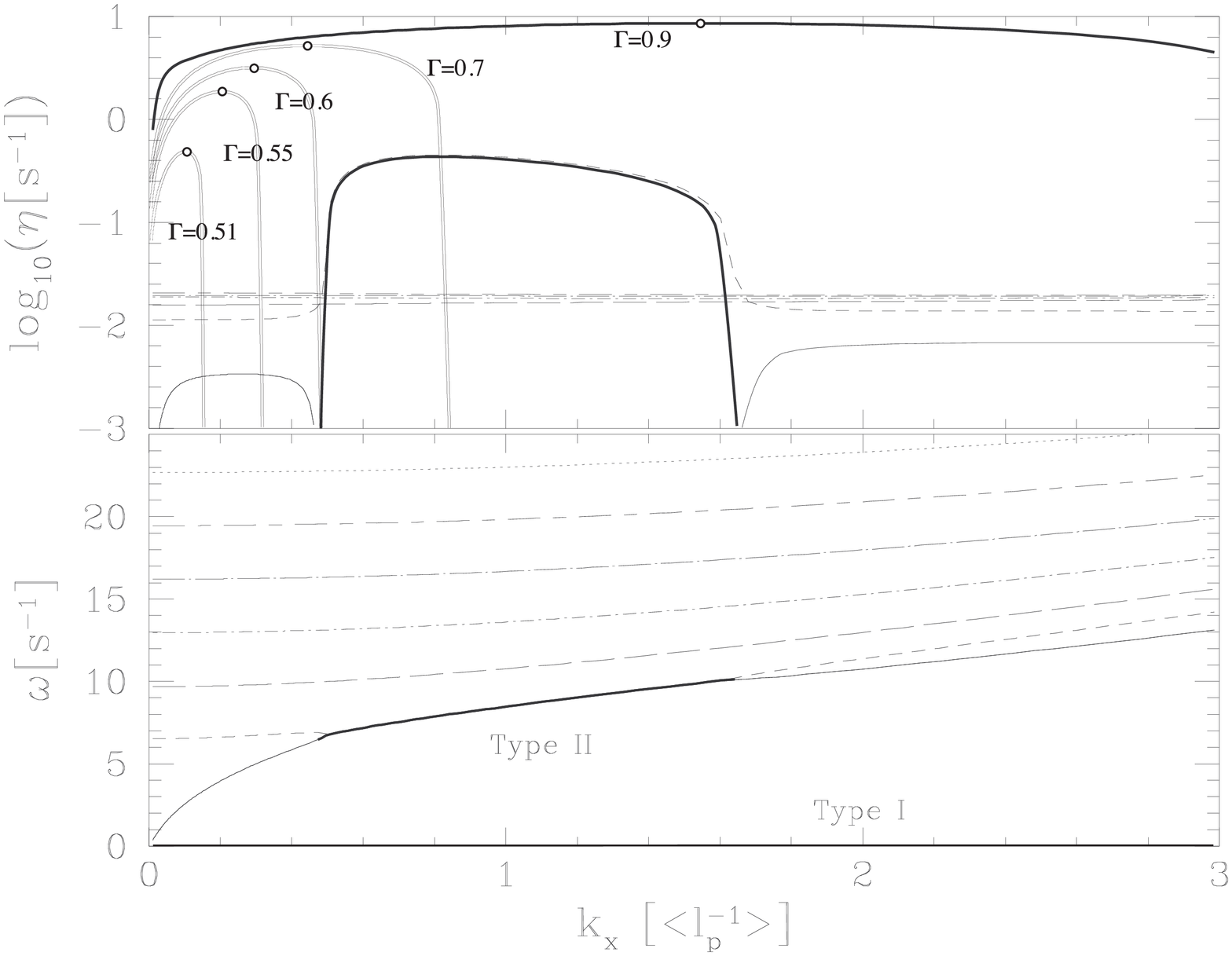,width=\figurewidth,angle=-90}}
\caption{The eigenvalues of the first atmosphere depicted in
Fig.~\ref{fig:atmos123} that has a fixed temperature imposed at its
bottom, as a function of the horizontal wavenumber $k_x$ in units of
the pressure scale height. The bottom panel is the real part of the
eigenmode while the top panel is the $\log_{10}$ of the absolute of the
imaginary part. Unstable modes are denoted by thick lines.  This
atmosphere exhibits both Type I and Type II unstable modes.  Below the
critical horizontal wavenumber, the Type I unstable mode has a purely
imaginary eigenvalue.  Above the critical horizontal wavenumber, the
absolute value of the real part is much larger than the absolute value
of the imaginary part namely, there is a small damping term arising
from radiation diffusion.  The additional double lines represent the
imaginary part of the Type I instability when $\Gamma$ is decreased
from 0.9 to 0.7, 0.6, 0.55 and 0.51, with which we see that the most
unstable wavenumber moves to the long wavelength limit.  }
\label{fig:atmos_kT}
\end{figure*}

\begin{figure*}[tbh]
\centerline{\epsfig{file=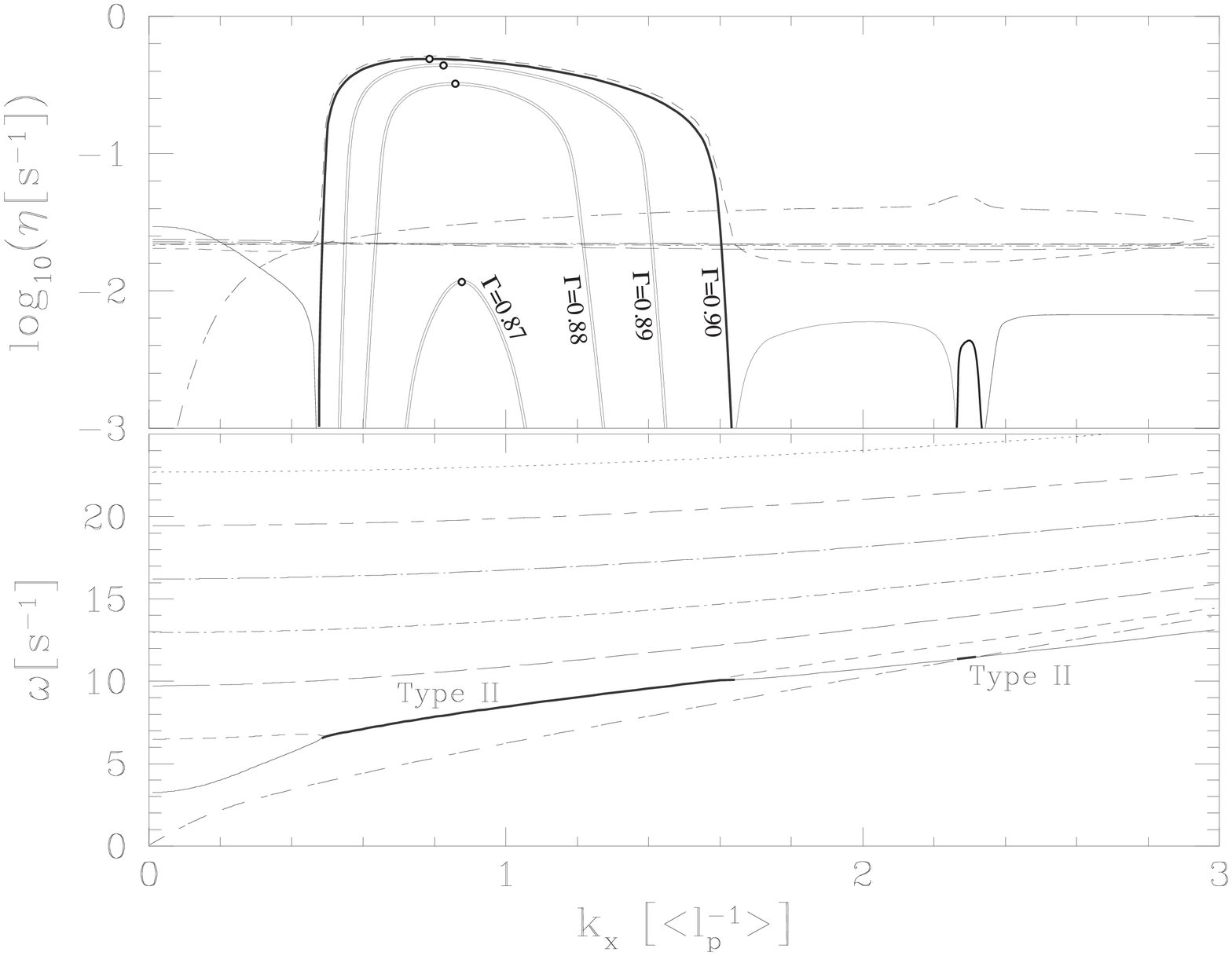,width=\figurewidth,angle=-90}}
 \caption{Same as the previous figure except that now the flux is kept
 constant as the bottom boundary condition. Here, the atmosphere is
 only unstable to the Type II instability. We see that the second
 eigen mode `merges' at two different horizontal wavelengths, with
 either the third eigenmode or the first eigenmode. The additional
 lines represent the imaginary part of the growth rate as $\Gamma$ is
 decreased. We see that the most unstable wavenumber changes only
 slightly with $\Gamma$. The growth rate is typically several times the
 oscillation period.}
\label{fig:atmos_kflux}
\end{figure*}

 Taking the two examples depicted in Fig.~\ref{fig:atmos123} with
 $\tau_{b}=10$, we now look at the locus of eigenmodes obtained as a
 function of $k_{x}$.  This is depicted in Fig.~\ref{fig:atmos_kT} for
 the fixed temperature case, which has both Type I and Type II
 unstable modes, and in Fig.~\ref{fig:atmos_kflux} for the fixed flux
 case, which has only Type II unstable modes.

 The Type I modes are found to be unstable for all $k_{x}$ smaller
 than a critical wavenumber.  This wavenumber decreases as $\Gamma$ is
 reduced.  That is to say, once the critical $\Gamma_\mr{crit}=1/2$
 for instability is surpassed, it is the longest wavelengths which
 first become unstable.  However, since $\Im(\omega)$ is found to be
 proportional to $k_{x}$ at long wavelengths, the most unstable mode
 has a finite horizontal wavelength. This is seen in the additional
 lines in Fig.~\ref{fig:atmos_kT} which are labeled with varying
 $\Gamma$'s. The additional lines describe the growth rate of the Type
 I unstable mode for different $\Gamma$'s with the most unstable
 horizontal wavelength marked with a small open circle.

 The Type II modes are seen to occur only in finite wavelengths.  In
 Fig.~\ref{fig:atmos_kflux}, we can actually see the second eigenmode
 become Type II unstable in one region of $k_{x}$ by merging with the
 third eigenmode and in another region, by merging with the first
 eigenmode.  As the $\Gamma$ is reduced towards the critical $\Gamma$
 needed for instability, we see that the most unstable mode does not
 vary by much.  The instability is triggered at a finite horizontal
 wavelength.  It is found to be triggered at $\Gamma_\mr{crit}\approx
 0.86$ and a wavenumber of $k_{x,\mr{crit}}\approx 0.88
 \left<l_{p}^{-1}\right>$.  These numbers are found to vary by a few
 percent if either the absorption opacity or the depth of the
 atmosphere are changed.
 
\subsubsection{Eigenmodes as a function of $\Gamma$}
\label{sec:modes_gamma}

 Last, we study the behavior of an atmosphere as a function of the
 Eddington parameter $\Gamma$.  We take the example depicted in
 Fig.~$\ref{fig:atmos_kT}$ and instead of changing $k_x$, we keep it
 fixed at $k_x=\left<l_\rho^{-1}\right>$ and change the Eddington
 parameter from $\Gamma=0.1$ to $\Gamma=0.98$.

\begin{figure*}[tbh]
\centerline{\epsfig{file=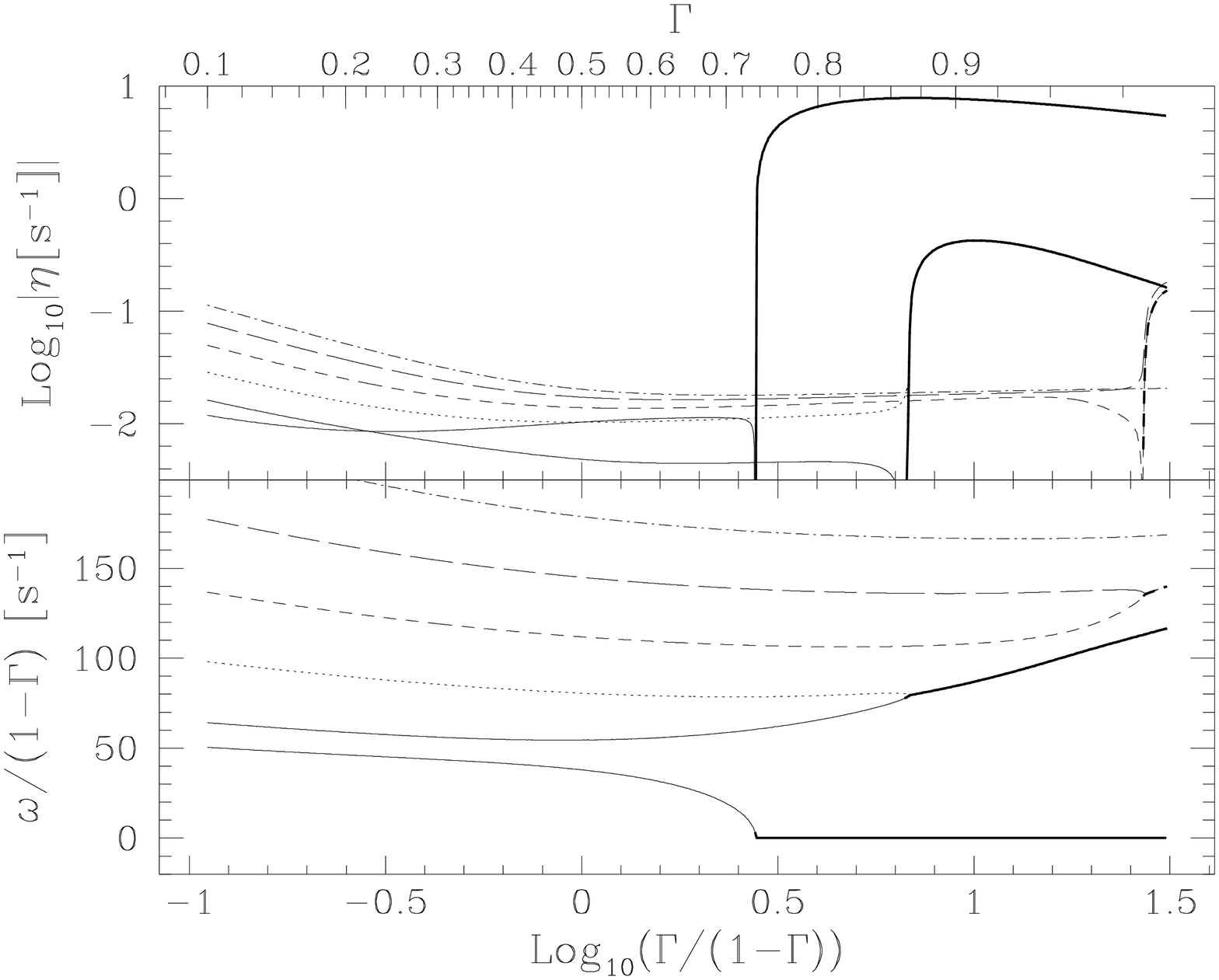,width=\figurewidth,angle=-90}}
\caption{ The eigenvalues of the first atmosphere described in
Fig.~\ref{fig:atmos123} (with a fixed $T$ at the bottom) as a function
of $\Gamma$, while $k_{x}$ is kept fixed at $\left< l_p^{-1}\right>$.
$\omega$ is multiplied by $1/(1-\Gamma)$ to normalize it to the
changing dynamical time scale as the atmosphere puffs up when $\Gamma$
is increased.  The graph is actually plotted as a function of
$\log_{10} (\Gamma/(1-\Gamma)) \approx \log_{10} \beta(\tau >> 1)$ in
order to extend the high $\Gamma$ region. Like in the previous
figures, unstable modes are depicted with heavy lines.  We see that as
the flux is increased, this particular atmosphere first becomes
unstable to the Type I instability (with a purely imaginary mode).  As
$\Gamma$ approaches the Eddington limit, more and more modes become
unstable to the Type II instability (which are mostly propagating).  }
\label{fig:GammaT}
\end{figure*}

 It is clear from Fig.~\ref{fig:GammaT} that Thomson scattering
 dominated atmospheres become unstable at large enough luminosities. 
 However, the critical luminosity is well below the Eddington limit. 
 They range from $0.5$ in the fixed temperature case to about $0.86$
 when a fixed flux is imposed at the bottom.
 
 The main feature apparent in the figure is that as the radiative flux
 increases, more and more modes become unstable to the Type II
 instability.  That is to say, it isn't only the lowest or second
 lowest vertical eigenmodes which are unstable.  The dynamic state
 that this atmosphere will reach obviously depends on the nonlinear
 behavior of the modes and their mutual interaction. It is clear
 however that it will not stay homogeneous close to the Eddington
 limit.

\section{The Type I Instability}
\label{sec:TypeI}

 The two main clues pointing to the origin of the Type I instability
 were seen in the previous sections. First, is was found that the
 instability arises only when the bottom boundary condition has a
 fixed temperature, as opposed to a fixed flux. Second, when it does
 arise, there is clearly an anti-correlation between the radiation
 energy density and the material density. We now proceed to show how
 this instability can arise with a toy model.
 
 We begin by showing how the boundary conditions affect the 
 correlation between gas and radiation. We will then proceed to show 
 how an anti-correlation results with the Type I instability.

\subsection{Non local effects arising from a global analysis}
\label{sec:boundaryeff}

 Since the boundary conditions were found to be crucial, we evidently 
 cannot study the instability with a local analysis. 
 We begin with an
 atmosphere in the NAR limit. Thus, at any given instant, the
 radiation field is governed by the equilibrium diffusion solution:
\be
\nabla \cdot {\bf F} = 0 ~~{\rm and}~~ {\bf F} = {c \nabla E \over
3 \chi \rho},
\ee
 and the boundary conditions imposed at the bottom and top of the
 atmosphere.

 To demonstrate how a positive or negative correlation can arise
 between $E_{1}$ and $\rho_{1}$, we explore a specific
 example. Suppose we have an unperturbed atmosphere in which
 $\chi_0=const$ that is perturbed by $\delta \chi_v = \delta
 \chi_v(x)$, i.e., that the perturbations are only a function of the
 horizontal direction $x$. We also assume for simplicity that the
 vertical variations in the atmosphere can be neglected at this point.
 This implies that the flux ${\bf F}$ and its perturbation do not
 depend on the vertical coordinate.

 The optical depth between the top and any given point in the
 atmosphere is
\begin{equation}
\tau = \int_z^{z_t} \chi_v dz = \tau_0 \left( 1+ \delta
\chi_v(x)/\chi_{v,0}\right),
\end{equation}
 where $\tau_0(z)$ is the unperturbed optical depth from infinity down
 to the mass element at $z$.

 Next, we need to impose boundary conditions. If we impose $\tau=0$ at
 the top of the atmosphere, then from the Eddington approximation we
 have that at the top $E_t = 2 F_{t} / c$. At the bottom we can impose
 a fixed temperature and therefore a fixed $E_b$, or, we can impose a
 fixed flux $F_{b}$.  An example for the implementation of a fixed
 temperature (or fixed $E_{b}$) at the bottom is the case of a very
 good energy conducting layer below the layer under consideration.

 The fixed flux boundary condition at the bottom corresponds to the
 case when the layer below the one under consideration has a very
 large heat transfer resistivity so that the gradient in the layer
 sets a fixed flux that cannot be changed by the atmosphere exhibiting
 the instabilities.

 When the temperature and energy density $E_b$ are chosen as fixed on
 the lower boundary, the flux $F$ and the energy density at any height
 are found to be:
\begin{eqnarray}
 F&=& {c E_b \over 2+ 3 \tau_b} ~~ = ~~const \nonumber
\\
 E&=& \left( 2+3 \tau \over 2+3 \tau_b \right) E_b ~~= ~~ \left( 2+3
 \tau \left( 1+\delta\chi_v(x)/\chi_{v,0} \right) \over 2+3 \tau_b
 \right) E_b
\nonumber \\
 & \approx & E_0 (\tau_0) \left( 1 - {6(\tau_b - \tau_0) \over (2+3
\tau_b) (2+3 \tau_0) } { \delta \chi_v(x)\over \chi_{v,0}}\right),
\end{eqnarray}
 where $E_0(\tau_0)$ is the radiation energy density at the point
 where the unperturbed optical depth was $\tau_0$.  This result
 implies that under a positive perturbation to the opacity, the change
 $\delta E(\tau_{0})$ in the radiation energy density will be
 negative:
\begin{equation}
 {\delta E (\tau_0) \over E} \approx - {6(\tau_b - \tau_0) \over (2+3
 \tau_b) (2+3 \tau_0) } { \delta \chi_v(x)\over \chi_{v,0}};
\end{equation}
 that is to say, there will be an anti-correlation between $\delta E$
 and $\delta\chi_v$. The coefficient of the anti-correlation isn't
 large. The largest anti-correlation is obtained if the lower boundary
 condition is fixed to be at $\tau_b = 2\sqrt{2}/3 \approx 0.94$ in
 which case $\delta E / E = -(3-2\sqrt{2}) \delta \chi_v / \chi_{v,0}
 \approx 0.172 \delta \chi_v / \chi_{v,0}$. The anti-correlation is
 larger than 0.1 for $5 \gtrsim \tau_b \gtrsim 0.2$.

 A negative correlation implies that the radiation pressure is
 anti-correlated with the density.

 When a fixed flux is imposed at the bottom, we have
\begin{eqnarray}
 F & = & F_0 = const \nonumber
\\
  E & = & (2 + 3 \tau) {F_0 \over c} \approx E_0 (\tau_0) \left( 1+ {3
  \tau_0 \over 2+3 \tau_0} { \delta \chi_v(x)\over \chi_{v,0}}
  \right).
\end{eqnarray}
 Here the correlation is positive, namely:
\begin{equation}
 {\delta E(\tau_0) \over E} \approx {3 \tau_0 \over 2 + 3 \tau_0} {
 \delta \chi_v(x)\over \chi_{v,0}}.
\end{equation}
 The coefficient of the correlation can be in this case larger than in
 the previous one and it increases with the optical thickness of the
 atmosphere. For $\tau_b \rightarrow \infty$, we have ${\delta E / E} =
 {\delta \chi_v(x)/ \chi_{v,0}}$.

 We conclude that under {\em the same local conditions} one can obtain
 qualitatively different behaviors according to the {\em boundary
 conditions} imposed. Should we have evaluated this atmosphere
 using a local analysis, we would have obtained that $\left<\delta
 p_\mr{rad}\right>=0$ since $k_z=0$. Clearly, the boundary conditions are
 important to the question of stability.

In a real atmosphere with vertically dependent unperturbed state and
modes, we could expect to see eigenmodes with both types of
correlations.  We point that the correlations do not depend on $\beta$
and hence such phenomena can occur in low luminosity atmospheres.  A
typical example is the formation of grains in atmospheres.  The
appearance of grains causes large opacity changes.

\subsection{Toy model with non local effects }
\label{sec:Toy}

 We assume for simplicity that the relevant perturbations (of the
 order of the scale height of the atmosphere) are in the NAR limit.
 We also assume that the gas is held within a narrow slab without any
 $z$ dependence. That is to say, we assume that the perturbation to
 the radiation density is proportional to the density perturbations:
\begin{equation}
 E_1 = \varepsilon E_0 {\delta \rho\over \rho}{~~\rm or
 ~equivalently}~~ p_\mr{rad,1} = \varepsilon p_\mr{rad,0} {\delta \rho \over
 \rho},
\end{equation}
 where $\varepsilon$ is the `efficiency' with which the average
 radiation energy density or pressure changes with changes in the
 opacity (which is a function of $\rho$). In \S\ref{sec:boundaryeff},
 $\epsilon$ was calculated for a very simplified case.  In reality,
 $\delta p_\mr{rad}$ is not simply proportional to $\delta \rho$ but is
 also a function of height, we just assume it to be so (this is why it
 is a `Toy' model!). We therefore leave $\varepsilon$ as an
 effective free parameter.

The equations describing the gas are the continuity and momentum
equations. The linearized versions of which are
\begin{eqnarray}
   {\partial \rho_1 \over \partial t} &=& - \nabla \rho_0 \cdot {\bf
v}_1
   - (\nabla \cdot {\bf v}_1) \rho_0 \\
   \rho_0 {\partial {{\bf v}_1}  \over  \partial t} &=
   & -\nabla p_1 - \nabla p_\mr{rad,1}.
\end{eqnarray}
Since the gas is in the NAR limit, its pressure perturbation is given
by
\be
   p_1 = c_T^2 \rho_1 + c_T^2 \rho_0 {E_1 \over 4 E_0} = \left( 1+
   {1\over 4}\varepsilon \right) c_T^2 \rho_1.
\ee
 The radiation pressure perturbation can be written as
\be
p_\mr{rad,1} = \beta p_{0} {\rho_1 \over \rho_0} = \beta c_T^2 \rho_1.
\ee
   Thus,
\begin{equation}
   {\partial^2 \rho_1 \over \partial t^2} - \Delta \rho_1 \left(1+
   \varepsilon/4 + {\varepsilon \beta} \right) c_T^2~=~0.
\end{equation}
 The first term in the parenthesis arises from the isothermal speed of
 sound and describes simple acoustic waves.  The second term arises
 because the NAR limit is not necessarily the isothermal
 limit. Consequently, the residual correlation between $\delta T$ and
 $\delta \rho$ imply that the relevant sound speed is modified to:
\be
 c_{\mr{NAR}}^2 \equiv c_T^2 (1+ \varepsilon/4).
\ee
The third term within the parenthesis arises from the work that the
radiation does in synchronization with the wave.

The corresponding dispersion relation is
\begin{equation}
\omega^2 = k^2 \left( 1+ \varepsilon(\beta+ 1/4)  \right)
c_T^2 \equiv k^2 c_{\mr{eff}}^2.
\end{equation}
 A negative $\varepsilon$ implies a decrease in the effective speed of
 sound. When the anti-correlation is large enough, the speed of sound
 $c_{\mr{eff}}^2$ becomes imaginary and the waves simply amplify
 without propagating. In most astrophysical systems, this cannot arise
 unless $\beta$ is sufficiently large because $\varepsilon$ is seldom
 large by itself\footnote{However, the following are two known
 exceptions.  (i) A photon + gas fluid with $\nu$'s playing the role
 of the radiation, (ii) An atmosphere with grains in which the opacity
 changes can be extremely large.}.

 If $\varepsilon$ is negative, the system will necessarily become
 unstable at some $L=L_{\mr{crit}}<L_{\mr{Edd}}$ for which $\beta$ is
 large enough.  Using the toy model and a typical value below the
 photosphere of $\varepsilon\approx -0.15$ which we should expect to
 see in an atmosphere with a $\tau_{b}$ of about 5, we have that the
 critical $\beta$ in this case is $\beta_{\mr{crit}}\approx
 1/\varepsilon - 1/4 \approx 6.5$, and the critical Eddington
 parameter is $\Gamma_{\mr{crit}}\approx \beta_{\mr{crit}}/
 (\beta_{\mr{crit}}+1) \approx 0.86$. The full numerical calculation
 gave $\Gamma_\mr{crit}=1/2$ for a deeper atmosphere. This discrepancy
 can actually be reconciled exactly with a more elaborate toy
 model. The feature missing in this simple model is the `feedback'
 that the change in the radiation flux has on the vertical extent of
 the atmosphere. Specifically, a lower density region will have a
 higher flux going through it and therefore a smaller effective
 gravity. Consequently, it would puff up. At a given height above the
 bottom, the radiation density will therefore be even higher because
 the Lagrangian coordinate will be lower. This will introduce a
 further anti-correlation between $\rho$ and $E$. This missing feature
 in the toy model also allows very deep atmospheres to be
 unstable. Without it, we saw in \S\ref{sec:boundaryeff} that as
 $\tau_{b}$ is increased, the anti-correlation tends to zero.

 Once an atmosphere is unstable to Type I instabilities, it would tend
 to open `chimneys' through which it is easier for the radiation to
 escape and also to accelerate material.


\section{The Type II Instability}
\label{sec:TypeII}

To understand the Type II instability, it is best to compare it to the
instability of radial $s$-modes (e.g., \citenp{G94}, and
\citenp{Pap97}). As we shall see, there are many similarities and a
few particular differences which will elucidate the origin of this
instability.

The main similarities between $s$-modes and the Type I instability 
are:
\begin{enumerate}
\rightskip 0pt
    \item The instabilities result with eigen-frequencies that describe
    modes with $e$-folding growth times of order the oscillation
    period. Each mode has a conjugate that is damped at the same
    rate. Moreover, as a control parameter is changed (e.g., $T_\mr{eff}$
    in $s$-modes or $\Gamma$ here), the conjugate pair arises when two
    modes with different real frequencies merge together.
    
    \item Both types of instabilities appear only in systems in which
    the radiation pressure dominates.
    
    \item Both instabilities are quite different from various `slow
    instabilities' such as the $\kappa$-mechanism and other Carnot 
    types.
    
    \item Both instabilities disappear when the system is driven out
    of the two fluid limit.  Namely, both need the radiation field to
    be described by the diffusion equations instead of being highly
    coupled to the gas.
    
\end{enumerate}

Nevertheless, the are a few critical differences between the two:

\begin{enumerate}
\rightskip 0pt    
    \item The Type II instability operates in Thomson atmospheres, 
    $s$-modes need a particular opacity law for them to be 
    unstable. In particular, they need $\bar{\kappa}_{T}>0$ which is 
    obtained in ionization zones. Thomson atmospheres are much more 
    general.
    
    \item Unstable $s$-modes are a radial phenomenon. The Type II is 
    intrinsically non-radial. Type II unstable modes do 
    not exist if the horizontal wavelength is significantly larger 
    than the pressure scale height of the system. It also implies that 
    it will be a local phenomenon in stars since the scale height is 
    usually much smaller than the radius.
    
    \item $s$-modes are localized to the top part of the atmosphere,
    where the NAR limit is obtained, by having a region with an
    effective speed $c_{\kappa}^{2}<0$.  The Type II unstable modes
    will be localized to the top because they are a very high $\ell$
    phenomenon.
    
    \item Last, unstable $s$-modes require that the gas temperature
    be given by the radiation temperature, namely, the NAR limit.
    Type II unstable modes where found to exist also when the
    atmosphere was switched to the adiabatic limit (though still in
    the two fluid limit) by reducing the absorptive opacity enough to
    have $\tau_\mr{cool} \gg \tau_\mr{dyn}$.
    
\end{enumerate}

We see that, mathematically, the origin of the two different
instabilities is similar.  The instabilities arise when the linearized
equations become non self-adjoint (e.g., \citenp{Pap97}) as it is this
characteristic that results in the complex conjugate pair of
eigenmodes.  In the case of $s$-modes, it was shown that it is because
the dispersion equation becomes third instead of second order.
Without the interaction with the diffusive radiation field, the
pressure and density have a local algebraic relation of the form
$p_{1} = v_{s}^{2} \rho_{1}$.  As such, the dispersion equation for
$\omega^{2}$ is self-adjoint and no complex roots are found for
$\omega^{2}$.  However, because the temperature is set by the
radiation field, the relation between density perturbation and total
pressure perturbation becomes differential (\citenp{G94}). To see
this, we look at the vertical component of eq.~(\ref{eq:radF}), and
perturb it:
\begin{equation}
    {\partial p_\mr{tot,1} \over \partial z} \approx {\partial p_\mr{rad,1}
    \over \partial z} = {1\over 3} {\partial E_{1} \over \partial 
    z} = - {\chi_{v,1} F_{z,0}\over c} - {\chi_{v,0} F_{z,1} \over c}
    \label{eq:radFper}
\end{equation}
In the case of radial perturbations, $F_{z,1}$ vanishes. Thus, in
$s$-modes, it is the first term on the r.h.s from which a differential
relation is obtained between $p_\mr{tot,1}$ and $\rho_1$ which
$\chi_{v,1}$ is a function of. The dispersion relation becomes third
order and $\omega^{2}$ can obtain complex conjugate roots.  The
differential nature of the pressure density relation is obtained by
perturbing the radiation transport equation under a fixed flux, as it
is in the two fluid limit (\citenp{G94}).  Because the variable
opacity, which is a function of the gas temperature, is one of the key
ingredients needed to get the differential relation, it is clear why
$s$-modes do not exist in the two fluid adiabatic limit.  In this
limit, the temperature of the gas, and therefore the opacity, is not a
function of the radiation field but of the gas alone.

The physical origin of the Type II instability is similar to that of
$s$-modes though not the same.  The different characteristics of
the two instabilities should point to the differences in
physical origin.

First, like $s$-modes, the Type II doesn't exist in the one fluid
limit and the radiation pressure must dominate, thus, it is probably
the diffusive nature of the radiation field which promotes the Type II
instability.  Second, since Type II exists also in Thomson
atmospheres, the form of the opacity is obviously not crucial.  It
also explains why the Type II exists also in the two fluid adiabatic
limit -- the instability is not sensitive to the properties of the
gas, whether its temperature is set adiabatically or by the radiation.
Thus, we hypothesize that the origin of the Type II instability is in
the differential nature that the relation between the radiation
pressure perturbation and the density perturbations obtains due to
non-radial perturbations.  The second term in eq.~(\ref{eq:radFper})
fits the required features.  It will not vanish if $F_{z,1}\neq 0$,
namely, if there are perturbations to the flux in the vertical
direction.  Since $\nabla \cdot {\bf F} = 0$ in the diffusive limit,
such a perturbation to ${\bf F}$ can arise only if there are
horizontal perturbations $F_{x,1}$ as well.  These arise only when
there are horizontal perturbations to the density on the same scale as
the vertical perturbations, hence the need for non-radial modes.

To check this hypothesis, we artificially modify in the radiation
equations the term responsible for {\em horizontal} fluxes.  Namely, we
write $\nabla E \rightarrow (\partial E / \partial z)\hz + u_{h} (\partial E
/\partial x) \hx $, where $u_{h}$ is an artificial factor.  For $u_{h}=1$ we
recover the correct diffusion equations.  For $u_{h}\rightarrow 0$ we
inhibit the flow of radiation in the horizontal direction.  If our
hypothesis is correct, then under the latter limit, the Type II
instabilities should disappear.  What we find in the numerical 
calculation is just that.

To summarize, the Type II instabilities originate from a differential
relation between the total pressure and the gas density. This
transforms the dispersion equations to a higher order equation in
$\omega$ that can have complex conjugate roots. The differential
`equation of state' needed to relate the density and total pressure
perturbations in $s$-modes, arises when perturbing the constant flux
equations and it is sensitive to the opacity. The required
differential relation in Type II unstable modes arises when horizontal
perturbations are taken into account, allowing horizontal fluxes.

\begin{figure*}[tbh]
\centerline{\epsfig{file=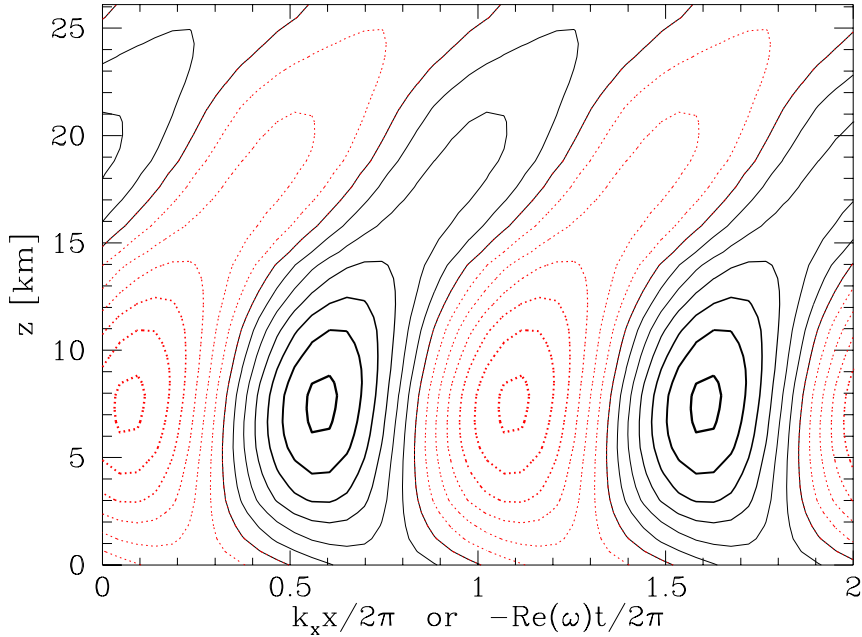,width=\figurewidth,angle=-90}}
\caption{ The radiation energy density of the Type II unstable mode of
the 2$^{\rm nd}$ atmosphere of Fig.~\ref{fig:atmos123} as a function of $z$
and either $x$ or $-t$.  If we choose the horizontal axis to be $x$,
we see a snapshot of the eigenmode.  Solid lines are positive
iso-contours of the radiation energy density while dotted are
negative.  The normalization is meaningless in the linear problem.
With time, this pattern moves to the right and amplifies according to
$exp(\eta t)$.  If we look at the evolution of the profile along a
vertical cut, its evolution of $\FE(z,t)$ without the exponential
growth will be given by the same plot with the horizontal axis being
$-t$, namely, it would appear as if structure is moving {\em
downwards}.}
\label{fig:bubbles}
\end{figure*}

How does a Type II unstable system look like?  To see this, we plot in
Fig.~\ref{fig:bubbles} a snapshot of the eigenmode of the system. 
Plotted is the energy density of the radiation as a function of $x$ or
$-t$.  The solution is proportional to $E_{1}(z)\exp (i k_{x}x - i
\Re(\omega) t) \exp(\Im(\omega)t)$.  Thus, if we fix $t$, we obtain
$E_{1}$ which varies with $z$ and is harmonic in the $x$ direction. 
If we vary the snapshot with time, we will find that it propagates to
the right side.  If we fix $x$, let $t$ and $z$ vary, and exclude the
exponential growth, we will obtain the same figure, with the
horizontal axis for the time flipped.  Namely, we will see at a fixed $x$,
structure moving {\em downwards}.  This is opposite to conventional
photon bubble picture.

\section{Discussion}
\label{sec:discussion}
\label{sec:nonlinear}

 We have seen that atmospheres with a significant radiative support
 always become unstable if the radiative pressure is increased
 sufficiently and becomes dominant. Both instabilities found are
 intrinsically non radial. Table \ref{table:instabilities} summarizes
 some of the characteristics of the two instabilities found and
 emphasizes the differences between them and other known instabilities
 -- convection and $s$-modes which are both dynamic instabilities, and
 the $\kappa$-mechanism which is a Carnot type instability.

\begin{table*}[tbh]

 \caption{Comparison between the instabilities found here and three
 others.}

{\small
\begin{center}
\begin{tabular}{| l | c c c c c |}
\hline
Instability:                                & Type I & Type II & Convection & $s$-mode &  $\kappa$-mechanism \\
\hline
Radial (exists for $k_x\rightarrow 0$)      &   -    &   -     &    -       &     +    &        +           \\
Propagating ($\Re(\omega)\neq 0$):          &   -    &   +     &    -       &     +    &        +           \\
Dynamic time scales ($\Im(\omega)\sim{\cal O}(1/\tau_\mr{dyn})$) 
                                            &   +    &   +     &    +       &     +    &        -           \\
Exists in adiabatic limit ($\tau_\mr{cool} / \tau_\mr{dyn} \rightarrow \infty $)
                                            &   +    &   +     &    +       &     -    &    ${^a}$  \\
Exists in NAR limit  ($\tau_\mr{cool} / \tau_\mr{dyn} \rightarrow 0 $)
                                            &   +    &   +     &    -       &     +    &    ${^a}$   \\
Exists in diffusion limit  ($\tau_\mr{diff} / \tau_\mr{dyn} \rightarrow 0 $) 
                                            &   +    &   +     &    +       &     +    &    ${^a}$   \\
Exists in one fluid limit  ($\tau_\mr{diff} / \tau_\mr{dyn} \rightarrow \infty $) 
                                            &   -    &   -     &    +       &     -    &    ${^a}$    \\
Requires large $p_\mr{rad}/p_\mr{gas}$      &   +    &   +     &    -       &     +    &        -          \\
Requires special opacity laws               &   -    &   -     &    -       &     +    &        +           \\
Significantly affected by boundaries        &   +    &   -     &    -       &     -    &        -           \\
\hline
\end{tabular}
\end{center}

${^a}$ The $\kappa$-mechanism per se requires that the gas will
{\em not} be fully in the adiabatic limit. Other similar Carnot type
instabilities can exist when the gas is {\em not} fully in the other
three limits. Namely, Carnot type instabilities can potentially exist
if the gas is not in one of the four corners of fig.~1.}
\label{table:instabilities}
\end{table*}

 Using a Toy model, we can understand the Type I instability. The
 anti-correlation between gas density and total pressure, dominated by
 the radiation, reduces the effective speed of sound. If the reduction
 is large enough, the speed of sound squared becomes negative and we
 have a purely imaginary frequency. This explains why $\Im(\omega)$ was
 found to be proportional to $k_x$ at long wavelengths.

 We have also seen with the Toy model that in addition to the
 radiative pressure term, changes to the temperature can contribute
 towards an instability. In Thomson atmospheres, this term is never
 large enough. However, when opacity variations can be very large, as
 is the case with dust formation, the temperature perturbation can by
 itself make the speed of sound become imaginary. In such a case, the lower
 limit for an instability is the lowest flux for which oscillations on
 the dynamical time scale are still not adiabatic.

 The `temperature' term contribution to the effective speed of sound
 can become important in a second type of cases in which the equation
 of state becomes significantly `softer' than that of an ideal
 gas. That is, when $\partial \ln p /\partial \ln T |_\rho \gg
 \partial \ln p /\partial \ln \rho |_T$. For example, in a system in
 which the fluid is composed of highly coupled radiation and plasma
 and the role of the radiation is played by a strong $\nu$ flux,
 then given temperature variations will now be more important in
 setting the effective speed of sound. When this happens, we can again
 expect to see an instability even in systems which are far from the
 Eddington flux (as long as oscillations on dynamical time scales are
 not adiabatic).

 The second type of instabilities found, the Type II, were found to be
 similar to $s$-modes.  The difference however is that Type II does
 not require special opacity laws on one hand, but on the other hand,
 it does require non radial oscillations.  The latter allows for a
 differential relation between the radiation pressure perturbations
 and the gas density which can result with a non self-adjoint
 dispersion equation for $\omega^2$.  Once triggered, the unstable
 modes describe highly distorted horizontally propagating waves as is
 depicted in Fig.~\ref{fig:bubbles}.  Interestingly, if a standing
 wave in the horizontal direction is composed out of two oppositely
 propagating waves, the result appears as `bubbles' propagating {\em
 downwards}!  That is, it appears as `anti-bubbles'.  The common
 conception that a `bubbly' phenomenon should arise as the Eddington
 limit is approached assumes that the `vacated' bubble through which
 it is easier for the radiation to escape can be held together by
 something (\citenp{Sp77}).  This is true in the case of `photon
 bubbles' in highly magnetized media where the material is forced to
 move only radially by vertical magnetic fields (\citenp{Aro}).  When
 no magnetic field is present, nothing is found to hold the material
 (\citenp{Sp77}), and as a result, these bubbles probably do not exit
 at all.  What we find instead is a very dynamic picture in which the
 structure or `phase speed' is actually downwards, such that the
 material in the `vacated' regions is constantly changing on dynamical
 time scales. This is the opposite of bubbles. Of course, this pattern
 movement downwards does not involve net motion of material, and it
 facilitates the transfer of radiation upwards.  What the exact form
 that this `anti-bubbles' will look like, when in the nonlinear regime,
 is of course an interesting open question that requires a full
 numerical simulation.

 A very important question which should be addressed in future work is
 the effects that the assumption of trapping of modes has. That is, it
 was assumed in the form of the boundary conditions, that acoustic
 waves are reflected at the boundaries above and below the layer. This
 need not necessarily be the case. If the atmosphere does not have a
 corona for example, where the temperature and speed of sound increase
 with height, any sound wave traveling through the atmosphere will
 just continue upwards and dissipate itself after forming shocks
 (which could in principle generate the required corona).  In any
 case, we expect both instabilities to be important because they
 amplify on a short dynamical time scale -- a mode does not need to
 travel back and forth to be amplified significantly. Moreover, the
 imaginary effective speed of sound in the Type I instability will be
 shown in a subsequent publication to be a signature of a phase
 transition in which the homogeneous state of the atmosphere is no
 longer the preferred or lowest energy state, that is, the instability
 found is just a manifestation of the system trying to get into a new
 equilibrium condition which is not the homogeneous one.

 Although a lengthy description of the nonlinear behavior is part of a
 forthcoming publication, a few words are in order about how such
 systems are expected to behave once these instabilities start to set
 in.

The results of \S\ref{sec:modes_gamma} have shown that either the Type
I or the Type II instabilities will necessarily set in before the
Eddington limit is reached.  Thus, as a luminous object approaches the
Eddington limit, convection will become important in the inner parts
of the object where the density is high enough and the cooling time
scale long (\citenp{JSO}), while the outer parts will be unstable to
the Type I or II instabilities where the two fluid limit is relevant,
even if the opacity is Thomson dominated.  Once either one of the
instabilities takes place, it is clear that the atmosphere will not
remain homogeneous.  If the nonlinear pattern that is eventually
formed propagates, as is the case in the Type II instability, each gas
element will experience periodic density variations.  While the peak
force experienced by each mass element may exceed the Eddington limit,
the time average force may be significantly below it \citep{Sha98}. 
In this case, an atmosphere can transfer through it a luminosity
greater than the Eddington luminosity without blowing itself apart. 
This will occur in the region of the atmosphere where perturbations
of order the scale height become optically thin.  From this region
upwards the effective opacity is again the microscopic one and the
luminosity becomes effectively super Eddington.  A predictable wind
will then be blown from that region upwards (\citenp{Novae}).

 In the case of the Type I instability, on the other hand, the pattern
 formed is stationary and it resembles that of `chimneys'.  The gas
 elements present in the lower density regions can witness a constant
 (in time) super-Eddington flux.  Although the average over the entire
 atmosphere can be a sub-Eddington luminosity, some gas elements in
 this case can be accelerated upwards through the chimneys, driving
 turbulence in the shear layers along the sides of the chimneys.
 However, this material will probably not be able to escape to
 infinity because it is unlikely that the escape speed will be
 attainable in the chimneys.  This will require strong shears that
 presumable dissipation by shocks will prevent it from occuring.
 Thus, once the Eddington flux is surpassed in the chimneys, it will
 likely generate `fountains' in which material accelerates upwards
 until the top of the chimneys is reached, then the material will fall
 back.  If the average luminosity is super Eddington, the material
 need not stagnate and it can be easily driven off to infinity as a
 wind.

 In both cases, the clear role that the classical Eddington limit
 plays in the homogeneous case is lost.  On one hand, a strong wind
 can exist already below the Eddington limit, driven for example b the
 energy dissipated from the unstable modes; while on the other hand,
 quasi-stationary configurations (with a wind) can exist above the
 Eddington limit.

 A more detailed treatment of the nonlinear behavior, which includes
 toy models as well as numeral simulations is underway in a
 forthcoming publication.


\section{Summary}
\label{sec:summary}
 
 The two main results found in this paper are the existence of two
 types of dynamically important instabilities that can take place in
 Thomson atmospheres.  The two instabilities exist in addition to the
 possibility of convection.  Following is a summary of the main
 results pertaining to these two important instabilities which will
 govern the behavior of atmospheres shining close to the Eddington
 limit.

\begin{enumerate}
\rightskip 0pt
        \item Under a linear analysis of a slab Thomson atmosphere,
        two dynamically important instabilities were found to exist
        for the first time.

        \item Both instabilities operate in the two fluid limit of
        optically thick atmospheres -- when the diffusive time scale
        is shorter than the dynamical time scale of the system. Hence,
        they operate from the photosphere down to where convection
        can become efficient.

        \item The Type I instability is discovered once a
        nonlocal analysis of the radiation field behavior is carried
        out. A fixed flux boundary condition quenches it off.

        \item The Type I instability does not propagate. It
        corresponds to the opening of `chimneys'. It appears in modes
        that have the radiation density on average anti-correlated
        with the gas density. The anti-correlation reduces the square
        of the effective speed of sound. When $c_{\mr{eff}}^{2}$ is
        negative, the instability appears and grows over a dynamical
        time scale.

       \item The Type I instability can take place also for $k_x
       \rightarrow 0$, however, the growth rate then tends to vanish:
       $\eta \propto k_x \rightarrow 0$.

       \item A second instability, the Type II, appears under more
       general boundary conditions, but at somewhat higher $\Gamma$'s
       than the Type I.

       \item The mathematical origin of the Type II instability is
       similar to that of $s$-modes. Since the hydrodynamic equations
       are coupled to a diffusion equation for the radiation, the
       relation between the pressure and density is differential. In
       the Type II instability, it arises from the term responsible for
       generating perpendicular radiative fluxes.

       \item The Type II instability describes horizontally
       propagating waves that amplify over dynamical time scale.  If a
       standing mode is composed out of two oppositely propagating
       modes, it has the appearance of `bubbles' moving downwards. 
       This arises because the phase velocity of the wave standing in
       both the horizontal and vertical direction is pointing down.  This
       is opposite to the intuitive `photon bubble' picture, in which
       the structure appears to be moving upwards.

       \item As the flux through an atmosphere increases towards the
       Eddington limit, Thomson scattering dominated atmospheres
       become unstable above a critical fraction of the Eddington
       flux. This instability is independent of and comes on top of
       the possibility of convective instability. The first mode to
       become unstable can be either Type I or Type II unstable. Type
       I appears at $\Gamma\gtrsim 0.5$. However, the exact value depends
       on the boundary conditions and thermal gradient. The Type II
       instability appears under more general boundary conditions but
       at $\Gamma \approx 0.85$.

       \item The most unstable horizontal wavelength for the type I
       instability is the longest wavelength in the system at
       $\Gamma_\mr{crit}$, but it becomes smaller as $\Gamma$ is
       increased. The most unstable mode for the type II instability
       is about $0.9 \left<l_p^{-1}\right>$. It doesn't change by much
       as $\Gamma$ is increased from $\Gamma_\mr{crit}$.

      \item The analysis assumed thus far that the modes can
      be trapped around the layer under consideration. However, since the
      two instabilities operate over a dynamical time scale, they are
      probably important even if the acoustic oscillations are not
      trapped, as is the case with convection for example.

     \item The unstable modes always tend to reduce the effective
     opacity of the medium (if the luminosity is sufficiently close to
     the Eddington luminosity), thereby increasing the effective
     Eddington luminosity. The radiation and the opacity must
     therefore be solved simultaneously and self
     consistently. Treatment of the inhomogeneities on a large scale
     (say scales larger than a few optical lengths) requires a new
     formulation of the equation of state because of the process of
     averaging over the inhomogeneities. This problem is deferred to a
     later paper.

\end{enumerate}

  All the results presented in this work assume that the mass opacity
  is constant. In general however, the opacity can be a function of
  the gas variables. This can be shown to not only change
  quantitatively the results but also to introduce the effects of
  additional instabilities, some of which have not yet been described.
  Moreover, it should be stressed that since the temperature of the
  matter and the radiation at any given point are not necessarily
  identical, non equilibrium effects cannot be neglected. These
  effects should be discussed in the future.

 Finally, if we are also to fully understand the behavior of the
 instabilities, we should analyze their nonlinear behavior. It is this
 non linear behavior that will set many of the observational
 properties of systems shining close to their Eddington limit.

\acknowledgments

 The author wishes to thank the comments and suggestions made by the
 editor, anonymous referee, Mitch Begelman and Roger Blandford as well
 as Caltech for the DuBridge Prize fellowship which supported him and
 CITA for the current support.


\section*{Appendix: A Third Instability?}

\newcommand{\ST}{S\&T~}
 The numerical linear analysis performed by \cite{ST99} (hereafter
 S\&T) is similar to the analysis carried out here. However, \ST
 report on one hand the finding of a third instability that is not
 found here. On the other hand, they do not find the two instabilities
 that are described in this work. This of course raises the following
 two questions. Why didn't they find the instabilities described here
 and why didn't the analysis described here result with their
 instability. We wish to address these two questions here.

 The answer to the first question is rather simple. \ST only analyze
 one type of boundary conditions. The particular boundary conditions
 they chose tend to promote a lowest eigenmode with a net positive
 correlation between the radiation pressure and the gas density. This
 can be seen in their eigenmode profiles by comparing the radiation
 energy density and the gas pressure. Thus, their boundary conditions
 do not allow the `Type I' instability to show up.

 The second instability, the `Type II', was found to operate on higher
 vertical eigenmodes than the lowest one. Since \ST used a relaxation
 method to solve the eigenvalue problem and not a shooting method as
 was carried out here, they were limited to the lowest eigenmode
 only. Thus, they couldn't find the second instability either.

 The instability found by \ST is similar to various `over-stability'
 instabilities, such as the $\kappa$-mechanism, in which the growth
 time scale is significantly longer than the oscillation period. In
 their case, they found growth rates of order 100's to 1000's of
 oscillation periods. Why wasn't this instability found here?

 We searched with our code the available parameter space and no
 instability, as the one described, was found. One should note that
 \ST themselves cite the unpublished work of \cite{Marzek} who studied
 the same problem but could not find this instability either. We
 could, however, artificially get an instability similar to the one
 described in S\&T, if we did one of the following:
\begin{enumerate}
\rightskip 0pt
\item Calculate the Doppler term wrong, or neglect it altogether.
\item Use an inaccurate unperturbed atmosphere. 
\item Have a mismatched opacity between the unperturbed atmosphere and the
perturbations.
\end{enumerate}

 Since \ST found in their analysis that the Doppler terms result with
 only a very minor correction to the eigenmodes, the analysis
 described in this work was incorrectly first executed without the
 Doppler terms.  It was then found that the finite speed of light
 terms -- those multiplied by $\fac$, result with an instability very
 similar to what was described by S\&T. When we added the Doppler
 terms, we found that the radiation drag that it induces is always
 comparable but larger than the instability arising from the $\fac$
 terms.  An analytical estimate for the size of the terms corroborates
 this and it shows that the absorption term, which introduces damping
 as well, can be either larger or smaller.  Namely, there are two
 possibilities.  In the first, \ST calculated the Doppler terms
 correctly and the origin of the instability in not with the $\fac$
 terms, in which case they must have used a very small $v_s/c$ ratio
 (found in a low temperature atmosphere) a fact which is not stated.
 The second possibility is that the instability does originate with
 the $\fac$ terms in which case they erroneously underestimated the
 Doppler term.  One should note that \ST use the Eulerian description
 of the radiation while we use the Lagrangian description, namely, we
 write the zeroth and first moments of the radiation field in the
 frame of reference of the moving material.  The former method is, as
 it turns out, significantly more complicated since it involves many
 terms instead of one.

 If the Doppler terms were calculated properly, a second way of
 artificially getting an instability is if the atmosphere used as the
 zeroth order, or unperturbed state, is inaccurate and doesn't solve
 the radiation equations accurately. \ST describe the exact
 unperturbed solution, which for the optically thin limits is isothermal
 and the thick limits polytropic. They then solve for the eigenmodes
 assuming one or the other and then in an atmosphere that has both
 regions. If the latter case is actually not the accurate solution but
 the two limits `glued' together, it would explain how an instability
 can arise. This is because the unperturbed state does not solve
 exactly the unperturbed equations. The modes display an instability
 `triggered' by the atmosphere's will to satisfy the hydrostatic
 equations. We have also found the same effect for the same reasons if
 the unperturbed atmosphere is calculated with only the scattering
 opacity (which dominates), while the linear analysis is calculated
 with the scattering and very small absorption opacity.

 Irrespective of what the resolution to this question is, even if the
 instability does really exist, it is dynamically less important than
 both the Type I and Type II instabilities.




\end{document}